\documentclass[]{pasj02} 

\usepackage{xcolor}

\jyear{2026}
\Received{}
\Accepted{}

\graphicspath{{../}} 

\begin{document} 

\title{ Mapping of the Cold Neutral Medium via H~\textsc{i} Phase Separation in an Atomic Cloud Undergoing Molecular Cloud Formation}

\author{
 Yamato \textsc{Matsuzuki},\altaffilmark{1}\orcid{0009-0007-2537-3611}\altemailmark\email{y.matsuzuki@a.phys.nagoya-u.ac.jp}
 Hiroaki \textsc{Yamamoto},\altaffilmark{1}\orcid{0000-0001-5792-3074}
 and
 Kengo \textsc{Tachihara},\altaffilmark{1}\orcid{0000-0002-1411-5410}
}
\altaffiltext{1}{Department of Physics, Graduate School of Science, Nagoya University, Furo-cho, Chikusa-ku, Nagoya, Aichi 464-8602, Japan}



\maketitle

\begin{abstract}

We investigate the atomic-to-molecular gas transition in the molecular cloud formation region HLCG 92$-$35 using GALFA-HI and NANTEN $^{12}$CO($J$=1--0) observations. The H\,\textsc{i} spectra were decomposed into Cold Neutral Medium (CNM), Lukewarm Neutral Medium (LNM), and Warm Neutral Medium (WNM) components with the ROHSA algorithm, and CNM structures were identified using Astrodendro. The derived H\,\textsc{i} mass ratio is CNM:LNM:WNM = 43:47:10, indicating that a substantial fraction of the gas resides in the intermediate-temperature phase. We identify 1792 CNM clumps with characteristic sizes of $\sim$1--2 pc and average densities of $\sim80~\mathrm{cm^{-3}}$, broadly consistent with theoretical predictions for CNM structures formed through thermal instability. The CNM mass spectrum exhibits a power-law slope similar to that of CO clumps, suggesting that molecular clouds inherit the hierarchical structure established in the atomic phase. Significant non-thermal linewidths and localized CNM--CO velocity offsets imply that the CNM consists of small-scale cloudlets with distinct kinematic properties. These results provide observational support for a scenario in which small-scale CNM structures produced by thermal instability and turbulent flows represent an early stage of molecular cloud formation.
\end{abstract}


\section{Introduction}
Star formation is a fundamental process driving the evolution of galaxies, and it originates in molecular clouds that form through the transition from neutral atomic hydrogen (H\,\textsc{i}) to molecular hydrogen (H$_2$). H\,\textsc{i} in the interstellar medium (ISM) is generally classified into three thermal phases: the high-temperature, low-density Warm Neutral Medium (WNM), the low-temperature, high-density Cold Neutral Medium (CNM), and the thermally unstable Lukewarm Neutral Medium (LNM), which occupies an intermediate regime between them (e.g., \cite{HeilesTroland2003b};\cite{Murray2018};\cite{McClure-Griffiths2023}). It is crucial to understand the physical processes that generate high-density, low-temperature structures, as this knowledge is key to revealing the mechanisms of molecular cloud formation and subsequent star formation. Therefore, it is essential to elucidate the multiphase structure of the H\,\textsc{i} gas and its evolution to understand the origins of star-forming regions.

The phase transition from H\,\textsc{i} to H$_2$ is expected to occur preferentially in denser and colder CNM. Numerous theoretical studies have explored the physical processes responsible for the formation of such CNM structures. \citet{Field1965} and \citet{Field1969} first introduced the concept of thermal instability, establishing the theoretical foundation for the two-phase model of atomic gas, in which H\,\textsc{i} exists as a mixture of the WNM and the CNM. (\cite{Wolfire1995}, \yearcite{Wolfire2003}) conducted detailed thermal equilibrium calculations that incorporated the major heating and cooling processes in the ISM, thereby quantitatively defining the temperature and pressure structure of the two-phase neutral gas. More recent theoretical studies suggest that CNM structures can form via turbulent condensation within their precursor WNM (with temperatures of $\sim 8000$ K and densities of $\sim 0.3~\mathrm{cm^{-3}}$). A number of numerical simulations (e.g., \cite{KoyamaInutsuka2000},\yearcite{KoyamaInutsuka2002}; \cite{HennebelleAudit2007}; \cite{InoueInutsuka2012}; \cite{Saury2014}; \cite{Kobayashi2022}) have shown that H\,\textsc{i} gas behaves as a turbulent, multiphase system in which CNM clumps with low volume-filling factors emerge as a consequence of local compressions in the WNM. The efficiency of CNM formation appears to depend on the properties of turbulence and shock compression in the WNM (e.g., \cite{Seifried2011}; \cite{Saury2014}; \cite{Bellomi2020}; \cite{Kobayashi2020}). Despite these advances, how H\,\textsc{i} gas, particularly the formation and evolution of CNM structures, sets the initial conditions for molecular cloud formation remains poorly understood.

Extensive H\,\textsc{i} absorption-line surveys have been conducted in the Milky Way in recent decades (e.g., \cite{HeilesTroland2003a}, \yearcite{HeilesTroland2003b}; \cite{Murray2015}, \yearcite{Murray2018}; \cite{Nguyen2024}). These surveys provide increasingly detailed statistical descriptions of the H\,\textsc{i} phases, enabling investigations of the physical properties of atomic hydrogen in the interstellar medium. The spin temperature of the CNM is typically $\sim 100$ K, and the CNM accounts for roughly 30\% of the total H\,\textsc{i} column density, with a significant fraction of thermally unstable gas also present. However, since absorption line studies rely on background continuum sources such as quasars, they probe only narrow lines of sight and cannot resolve the internal structure of CNM clouds on sub-parsec scales.

H\,\textsc{i} 21 cm emission line observations are essential for studying the spatial distribution of H\,\textsc{i} gas. Large-scale H\,\textsc{i} surveys of the Milky Way and nearby galaxies have been conducted (e.g., \cite{Kalberla2005}; \cite{HI4PI2016}; \cite{Peek2018}), and methods for quantifying the multiphase structure of H\,\textsc{i} emission have advanced considerably. Traditionally, Gaussian decomposition based on differences in line widths has been utilized to distinguish between the WNM and CNM components (e.g., \cite{KulkarniFich1985}; \cite{Verschuur1995}). More recently, semi-automated Gaussian decomposition techniques have been applied to large-scale surveys to derive H\,\textsc{i} phase fractions (\cite{HaudKalberla2007}; \cite{KalberlaHaud2018}). In this context, ROHSA, introduced by \citet{Marchal2019}, performs regularized multi-Gaussian decomposition by exploiting the spatial coherence between adjacent lines of sight. This approach has established itself as a powerful method for reconstructing physically consistent WNM and CNM distributions across wide fields of view (e.g., \cite{Marchal2021}; \cite{MarchalMiville-Deschenes2021}; \cite{Buckland-Willis2025}).


Molecular clouds at high galactic latitude are ideal laboratories for studying molecular cloud formation. Located at $|b| > 20^\circ$, these clouds are expected to lie a few hundred parsecs from the Sun, given that the CO gas scale height of the Galactic disk is $\sim$ 100 pc \citep{Dame1987}. They have relatively few overlapping background structures, simplifying the analysis. Furthermore, they are generally diffuse, and are often considered to represent early evolutionary stages in the transition from H\,\textsc{i} clouds to CO-bright molecular clouds.

\begin{figure}
 \begin{center}
  \includegraphics[width=\columnwidth]{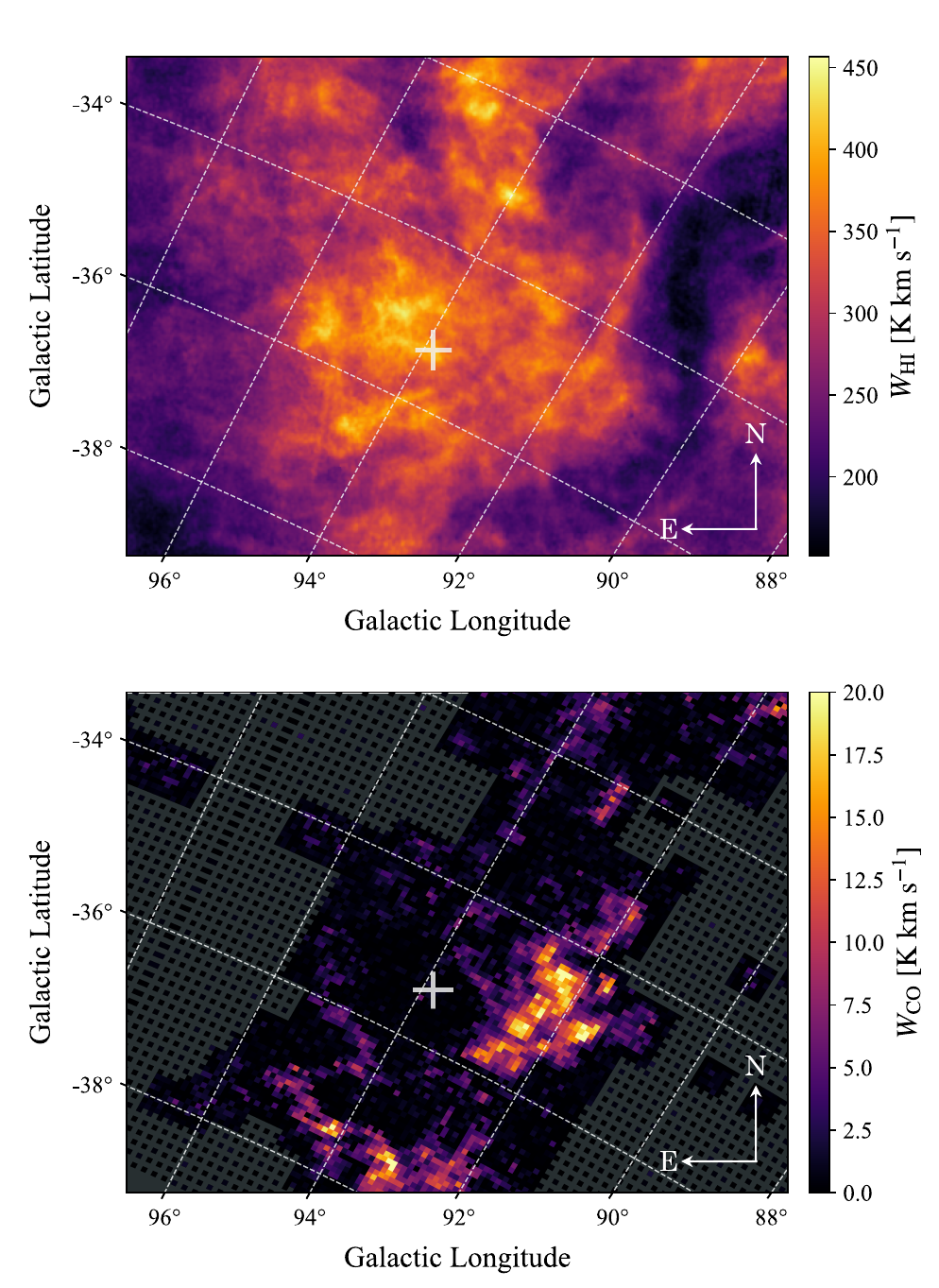} 
 \end{center}
\caption{
(Top) Integrated intensity map of the H\,\textsc{i} emission in the HLCG 92$-$35 region, integrated over the velocity range $-95.5~\mathrm{km~s^{-1}} \leq v_\mathrm{LSR} \leq 69.1~\mathrm{km~s^{-1}}$.
(Bottom) Integrated intensity map of the $^{12}$CO($J=1\text{--}0$) emission in the same region, integrated over $-20~\mathrm{km~s^{-1}} \leq v_\mathrm{LSR} \leq 10~\mathrm{km~s^{-1}}$. Gray indicates regions outside the observing area. The plus symbol marks the approximate center of HLCG 92$-$35. North and east are indicated by arrows.
{Alt text: Two Galactic coordinate maps of the HLCG 92$–$35 region. The top panel shows integrated neutral hydrogen intensity and the bottom panel shows integrated CO intensity. Unobserved areas are masked.}}\label{fig:fig1}
\end{figure}

HLCG 92$-$35, discovered by \citet{Yamamoto2003}, is located at $(l, b) = (92^\circ, -35^\circ)$ and is part of the high-latitude molecular cloud complex MBM 53, 54, 55 \citep{Magnani1985}. As shown in Figure~\ref{fig:fig1}, H\,\textsc{i} gas is abundant around the central region marked by the "+" symbol, whereas CO-emitting molecular clouds are predominantly distributed toward the western side of the cloud. The total molecular mass of HLCG 92$-$35 is estimated from $^{12}$CO observations to be $\sim330~M_\odot$. 
The total molecular mass of HLCG 92$-$35 is estimated from $^{12}$CO observations to be $\sim330~M_\odot$. The same study also found that HLCG 92$-$35 exhibits the largest far-infrared excess relative to the H\,\textsc{i} column density among the observed regions, suggesting the presence of substantial amounts of molecular hydrogen (H$_2$) not traced by CO. Furthermore, the ratio of far-infrared excess luminosity to $^{13}$CO mass is approximately five times higher than that of the surrounding MBM clouds.
These characteristics therefore suggest that HLCG 92$-$35 represents an early evolutionary stage of molecular cloud formation. The MBM 53-55 complex is thought to be part of an evolved supernova remnant, and the H\,\textsc{i} gas has been reported to be expanding at a velocity of about $18~\mathrm{km~s^{-1}}$ \citep{Gir1994}. This dynamical environment provides a plausible source of external compression, and the morphology of HLCG 92$-$35 is consistent with a scenario in which a shock wave from the west has recently triggered molecular cloud formation. Because this region lies at a relatively high Galactic latitude of $b=-35^\circ$, line-of-sight confusion from unrelated interstellar structures is expected to be significantly lower than along low-latitude sightlines, allowing a cleaner characterization of its physical properties.

In this study, we investigate the spatial distribution of H\,\textsc{i} phases and CNM clump structures in the high-latitude molecular cloud HLCG 92$-$35 to observationally elucidate the evolutionary process from the atomic to the molecular phase. We apply the multi-Gaussian decomposition method ROHSA to high-resolution H\,\textsc{i} data of the HLCG 92$-$35 region. By analyzing the phase separation of H\,\textsc{i} components, we aim to reveal how atomic gas evolves into molecular clouds. The structure of this paper is as follows. Section 2 describes the H\,\textsc{i} and CO data used in this study. Section 3 outlines the multi-Gaussian decomposition method ROHSA and the model selection process. Section 4 presents the results of the H\,\textsc{i} component separation in the HLCG 92$-$35 region. Section 5 discusses the spatial distributions and physical properties of the identified H\,\textsc{i} phases and their implications for the molecular cloud formation process.

\section{Data}
Figure \ref{fig:fig1}(a) and (b) show the integrated intensity maps of H\,\textsc{i} and $^{12}$CO ($J$~=~1--0) emission toward the HLCG 92$-$35 region. The distance to the MBM 53, 54, 55 complex, which encompasses HLCG 92$-$35, was estimated to be $\sim 150$ pc by \citet{Welty1989} based on stellar Na\,\emissiontype{I} absorption measurements. In this study, we adopt a distance of 150 pc for this region.

\subsection{H\,\textsc{i} data}
The H\,\textsc{i} data used in this study are taken from the GALFA-H I Data Release 2 (DR2) archive \citep{Peek2018}. The observations were carried out with the 305 m Arecibo radio telescope, providing an angular resolution of $4'$ on a $1'$ grid. At a distance of $150$ pc (the assumed distance to HLCG 92$-$35), this corresponds to a spatial resolution of $\sim 0.18$ pc. The velocity resolution is $0.184~\mathrm{km~s^{-1}}$, which is sufficient to resolve the narrow spectral features characteristic of the CNM.

Details of the GALFA-H I survey observations and data reduction procedures are described in \citet{Peek2018}. The analyzed field is shown in Figure \ref{fig:fig1}(a). The data cube was trimmed to a velocity range of $-95.5 \leq v_\mathrm{LSR} \leq 69.1 \mathrm{km~s^{-1}}$, resulting in a data size of 350 $\times$ 500 $\times$ 900 voxels (height $\times$ width $\times$ velocity channels). The H\,\textsc{i} emission is mainly concentrated in the range $-20  \leq v_\mathrm{LSR} \leq 10 ~\mathrm{km~s^{-1}}$, with no significant intermediate- or high-velocity components detected.
\subsection{CO data}
We used $^{12}$CO ($J$~=~1--0) data obtained with the NANTEN 4 m radio telescope to trace the molecular cloud. The survey covers an area of $\sim 141$ deg$^2$ within the Galactic coordinates $82^\circ \leq l \leq 98^\circ$ and $-45^\circ \leq b \leq -29^\circ$, encompassing the MBM 53, 54, and 55 complex. In this study, we utilized the portion of the dataset corresponding to the vicinity of HLCG 92$-$35.

The initial observations were carried out on a grid spacing of $8'\cos b$ in Galactic longitude and $8'$ in Galactic latitude. Regions exhibiting significant $^{12}$CO ($J$~=~1--0) emission were subsequently re-observed at a finer spacing of $4'\cos b$ and $4'$, respectively. The angular resolution of the telescope was $\sim$ $2'.6$, the velocity resolution was $0.10~\mathrm{km~s^{-1}}$, and the typical rms noise level was about $0.5$ K per channel. The observed velocity range was $-25  \leq v_\mathrm{LSR} \leq 10 ~\mathrm{km~s^{-1}}$, consistent with the velocities of the main H\,\textsc{i} emission components. Details of the observations and data reduction are given in \citet{Yamamoto2003}.

To ensure spatial consistency between the H\,\textsc{i} and CO datasets, the CO data cube was cropped to match the H\,\textsc{i} field and analyzed over the same area. The analyzed CO region is shown in Figure \ref{fig:fig1}(b).

\section{H\,\textsc{i} spectral decomposition method}

To decompose the H\,\textsc{i} spectra, we employed the multi-Gaussian decomposition algorithm ROHSA (Regularized Optimization for Hyper-Spectral Analysis; \cite{Marchal2019}). ROHSA performs spectral decomposition by fitting a linear combination of Gaussian functions to the observed H\,\textsc{i} profiles. The fitted model $T_\mathrm{model}$ is expressed as the sum of multiple Gaussian components:
\begin{equation}
    T_\mathrm{model}(\boldsymbol{r}, v_z)=\sum_{i=1}^NG_i(\boldsymbol{r}, v_z)
\end{equation}
where each Gaussian component at position $\boldsymbol{r}$ and line-of-sight velocity $v_z$ is defined as
\begin{equation}
    G_i(\boldsymbol{r}, v_z)=a_i(\boldsymbol{r})\exp{\left(-\frac{(v_z(\boldsymbol{r})-\mu_i(\boldsymbol{r}))^2}{2\sigma_i ^2(\boldsymbol{r})}\right)}
\end{equation}
where $a_i(\boldsymbol{r})$, $\mu_i(\boldsymbol{r})$, and $\sigma_i(\boldsymbol{r})$ represent the amplitude (= peak brightness temperature), central velocity, and velocity dispersion of the $i$-th Gaussian component, respectively.

In general, increasing the number of Gaussian components $N$ reduces the residuals and improves the fit to the observed spectra. However, using an excessively large number of components may result in overfitting to noise and does not necessarily yield a physically meaningful decomposition. To mitigate this, ROHSA adopts a regularized nonlinear least-squares approach that enforces spatial coherence among neighboring pixels.

This regularization allows for a consistent decomposition even in complex lines of sight where multiple components with different spatial distributions and central velocities overlap. Furthermore, ROHSA performs the optimization in a multi-resolution framework, in which the fitting is carried out progressively from low to high spatial resolution. This hierarchical approach ensures that both large-scale and small-scale structures are properly captured in the final model.

\subsection{ROHSA parameters}

When performing Gaussian decomposition with ROHSA, it is necessary to specify the number of Gaussian components $n_\mathrm{gauss}$, along with the strengths of the regularization terms: $\lambda_a$, $\lambda_\mu$, and $\lambda_\sigma$ to control the spatial smoothness of the amplitude, center velocity, and velocity dispersion maps, respectively, and $\lambda'_\sigma$ to minimize the variance of the velocity dispersion to facilitate multiphase separation (see \cite{Marchal2019} for more details on these parameters). Setting these hyperparameters to excessively large values may lead to unphysical or unnatural fits, as it oversmooths the parameter maps and erases genuine small-scale structures.These parameters must be optimized for each dataset individually. ROHSA allows the amplitude of a Gaussian component to be set to zero during the fitting process. Therefore, $n_\mathrm{gauss}$ represents the maximum number of Gaussian functions used for the decomposition, and it does not necessarily mean that all spectra in the region are fitted with exactly $n_\mathrm{gauss}$ components.

In this study, we searched the parameter space over $n_\mathrm{gauss}=6$--10, $(\lambda_a, \lambda_\mu, \lambda_\sigma) = 1$, 10, 100, 1000, and $\lambda'_\sigma = 1$, 10, 100. As in previous work (e.g., \cite{Taank2022}), we adopted identical values for $\lambda_a$, $\lambda_\mu$, and $\lambda_\sigma$. This assumption substantially reduces the number of free hyperparameters and makes the parameter search computationally tractable.

To evaluate how accurately the model derived by ROHSA reproduces the observed data, we introduced the ratio of the noise RMS to the residual RMS (hereafter referred to as the r.m.s. ratio), defined as $(\mathrm{noise~RMS})/(\mathrm{residual~RMS})$.

This quantity compares the noise level in the observed data with that in the residual data of the model: values close to unity indicate that the ROHSA model reproduces the data with high fidelity.

The noise RMS was calculated from emission-free regions (pixels 500--650 and 1350--1400), while the residual RMS was derived from the residuals in the emission-dominated region (pixels 650--1350).

This type of residual analysis for model evaluation has been used in previous studies(e.g., \cite{Taank2022}). While their evaluation was based on the $\chi^2$ statistic, the present study employs residual analysis as a complementary measure of model performance.

\section{Result}
\subsection{Model selection}

\begin{figure*}[t]
 \begin{center}
  \includegraphics[width=\textwidth]{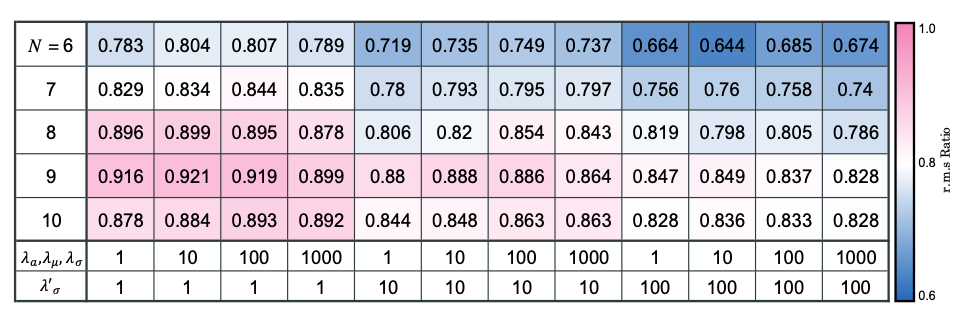}
 \end{center}
 \caption{r.m.s ratios for 60 different parameter combinations. The horizontal axis represents the regularization parameters $(\lambda_{\boldsymbol{a}}, \lambda_{\boldsymbol{\mu}}, \lambda_{\boldsymbol{\sigma}})$ and $\lambda_\sigma'$, while the vertical axis indicates the number of Gaussians used. r.m.s ratios close to 1 indicate higher fitting accuracy. In this study, the parameter set $(N, \lambda_a, \lambda_\mu, \lambda_\sigma, \lambda_\sigma') = (9, 10, 10, 10, 1)$ yielded the best performance.}
 \label{fig:fig1b}
\end{figure*}

Figure \ref{fig:fig1b} summarizes the results for a total of 60 parameter combinations. The indicator shown is the r.m.s. ratio (noise RMS / residual RMS). 

The r.m.s. ratio ranges from $0.64$ to $0.92$, exhibiting a general trend that it approaches unity as the number of Gaussian components $N$ increases and the constraint term $\lambda'_\sigma$ decreases.
However, when $N$ becomes excessively large, the r.m.s. ratio tends to decrease again.
While increasing $N$ allows the model to reproduce more complex spectral structures, it can also weaken the spatial regularization and the coherence among adjacent pixels, occasionally leading to a local degradation of the fitting accuracy. The fact that the r.m.s. ratio remains below unity for all fittings indicates that the residuals of the model exceed the observational noise level. This is likely due to limitations in spatial resolution, which prevent a complete reproduction of the observed spectra using a simple sum of Gaussian components, as well as to the presence of small residual structures such as baseline variations that are not captured by the model.
Among the tested combinations, the parameter set $(N, \lambda_a, \lambda_\mu, \lambda_\sigma, \lambda'_\sigma) = (9, 10, 10, 10, 1)$ yielded the best performance and was therefore adopted for the subsequent analysis.

Figure \ref{fig:fig2} (top) shows a scatter plot of all Gaussian components obtained using the selected parameter set $(N, \lambda_a, \lambda_\mu, \lambda_\sigma, \lambda'_\sigma) = (9, 10, 10, 10, 1)$. The horizontal axis represents the full width at half maximum (FWHM = $2\sqrt{2\ln2}\sigma$, where $\sigma$ is the velocity dispersion), and the vertical axis indicates the peak intensity. In total, 1,575,000 Gaussian components corresponding to all fitted spectra are plotted, including those with zero amplitude as allowed by ROHSA. The components are distributed over a range of $1$--$35~\mathrm{km~s^{-1}}$ in FWHM and up to $\sim$30 K in peak intensity.

The distribution in the FWHM--intensity plane can be separated into three distinct groups, which are hereafter referred to as clusters A, B, and C. Cluster A consists of narrow-line components, corresponding to Gaussian indices $G_2$, $G_3$, $G_4$, $G_5$, $G_6$, and $G_8$. Cluster B represents intermediate line widths ($G_1$), while cluster C corresponds to broad-line components ($G_7$, $G_9$). The Gaussian indices are arbitrary labels assigned by ROHSA and do not imply any physical ordering. The mean parameter values for each Gaussian component are summarized in Table \ref{table:table2}.

Cluster A is characterized by narrow line widths ($\mathrm{FWHM} \lesssim 12~\mathrm{km~s^{-1}}$), and is consistent with low-temperature gas within the H\,\textsc{i} medium. Previous all-sky H\,\textsc{i} Gaussian decomposition studies have shown that CNM components typically exhibit FWHM values in the range of  $\sim 0.8$--$13~\mathrm{km~s^{-1}}$ (see, e.g., Fig.~3 in \cite{KalberlaHaud2018}. The distribution of Cluster A is consistent with this range, and we therefore identify Cluster A as CNM in this work.

Cluster B, which corresponds to intermediate line widths, is often associated with thermally unstable gas, commonly referred to as the LNM or unstable neutral medium (UNM) (e.g., \cite{KalberlaHaud2018}; \cite{Marchal2021}). In the following, Cluster B is referred to as LNM. Although this component lies between the CNM and WNM in terms of characteristic line width (and hence temperature), it is not necessarily thermally unstable in all cases; rather, it likely represents gas in intermediate thermodynamic states.

Cluster C consists of broad-line, low-amplitude components, with peak intensities typically of only a few kelvin. Their broad linewidths and low peak intensities are characteristic of WNM-like components. We therefore refer to them as WNM components in this study, although their physical nature remains uncertain. Similar low-amplitude broad components have also been reported in other ROHSA-based studies (e.g., \cite{Marchal2021}; \cite{Taank2022}), suggesting that such features are a common outcome of Gaussian decomposition. 
Broad-linewidth Gaussian components may be affected by large-amplitude baseline ripples. To assess their impact, we estimated the baseline ripple amplitude from the increase in the variance of the brightness temperature, $T_{\mathrm{B}}$, with increasing velocity interval width ($30$--$300$ channels) within a line-free velocity range ($-180$ to $-124~\mathrm{km,s^{-1}}$), where the contribution from H\,\textsc{i} emission is expected to be negligible. The characteristic baseline ripple amplitude was estimated to be $\sim$ 0.087 K in the GALFA-HI data.Among the WNM Gaussian components identified in this study, approximately 86\% have peak amplitudes exceeding three times the estimated baseline ripple amplitude ($\sim~0.26$ K). We therefore expect that the influence of baseline ripples on the detection and parameter estimation of most WNM Gaussian components is limited.


Figure \ref{fig:fig2} (top) shows the distribution of Gaussian components in the FWHM–peak intensity plane for the CNM, LNM, and WNM phases. Figure \ref{fig:fig2} (bottom) presents the histogram of FWHM values for the Gaussian components associated with each phase. While the phase separation is generally clear, a slight overlap is observed between the CNM and LNM populations, as well as marginal overlap among the CNM, LNM, and WNM components.
Although a small overlap is visible between the CNM, LNM, and WNM populations in the FWHM–peak intensity plane (Figure \ref{fig:fig2}), the degree of cross-contamination is negligible. We quantified the contamination by counting the number of Gaussian components assigned to each phase within representative FWHM intervals, adopting boundaries of 12 and 25~$\mathrm{km~s^{-1}}$, which approximately correspond to the typical extents of the three populations in the FWHM distribution. Within the representative CNM interval (FWHM $< 12~\mathrm{km~s^{-1}}$), 99.87\% of the components are classified as CNM, while only 0.12\% and less than 0.01\% belong to the LNM and WNM populations, respectively. Within the representative LNM interval ($12$ $\leq$ FWHM $< 25~\mathrm{km~s^{-1}}$), 99.84\% of the components are classified as LNM, with contamination from CNM and WNM components of only 0.05\% and 0.10\%, respectively. Within the representative WNM interval (FWHM $\geq 25~\mathrm{km~s^{-1}}$), 99.99\% of the components are classified as WNM. Therefore, despite the slight overlap visible in Figure \ref{fig:fig2}, the three populations are well separated statistically, and the level of cross-contamination between phases is negligible.

Examples of the spectral fitting results are shown in Figure \ref{fig:fig3}. The observed H\,\textsc{i} spectra are plotted in gray, and the best-fit profiles obtained with ROHSA are shown in green. The individual Gaussian components corresponding to the CNM, LNM, and WNM phases are indicated by blue, purple, and red curves, respectively, consistent with the phase definitions in Figure \ref{fig:fig2}.

\subsection{Column density maps}
Figure \ref{fig:fig4} shows the spatial distributions of the column densities for the CNM, LNM, and WNM components defined in the previous section. The column density of each component was calculated by integrating the brightness temperature profile $G(v)$ of the corresponding Gaussian component obtained from the spectral decomposition:
\begin{equation}
    N_\mathrm{H\,\textsc{i}} = 1.822\times 10^{18}\int G(v)dv
    \quad [\mathrm{cm^{-2}}]
\end{equation}
where $G(v)$ represents the brightness temperature profile of each fitted Gaussian component. This calculation assumes that the H\,\textsc{i} gas is optically thin ($\tau \ll 1$). The total H\,\textsc{i} mass is given by
\begin{equation}
    M_\mathrm{H\,\textsc{i}} = \sum m_\mathrm{H}N_\mathrm{H\,\textsc{i}}\times S\quad [M_\odot]
\end{equation}
where $m_\mathrm{H}$ is the mass of a hydrogen atom ($1.6737\times10^{-27}~\mathrm{kg}$) and $S$ is the projected area of each pixel.

If the H\,\textsc{i} gas is optically thick, the column density should instead be estimated as
\begin{equation}
    N_\mathrm{thick, H\,\textsc{i}} = 1.822\times 10^{18}\int \frac{\tau(v)G(v)}{(1-e^{-\tau(v)})}dv\quad [\mathrm{cm^{-2}}]
\end{equation}
for which $N_\mathrm{H\,\textsc{i}}$ remains below $N_\mathrm{thick, H\,\textsc{i}}$.


However, accurate correction for optical-depth effects necessitates independent estimates of both the spin temperature and the optical depth, which cannot be constrained by emission-line data alone. To date, the optically thick column density $N_\mathrm{thick, H\,\textsc{i}}$ has been derived from combined on–off spectra toward radio continuum background sources, and the corresponding statistical correction factor $f = N_\mathrm{thick, H\,\textsc{i}}/N_\mathrm{H\,\textsc{i}}$ has been investigated. Although this correction factor varies among different regions, observations at high Galactic latitudes suggest that for column densities of $\sim10^{20}\,\mathrm{cm^{-2}}$, $f$ is close to unity, and even at $\sim10^{21}\,\mathrm{cm^{-2}}$, $f \sim 1.3$ (e.g., \cite{Lee2015}; \cite{Nguyen2019}, \yearcite{Nguyen2024}). Even if the column densities derived in this work are underestimated by up to $\sim$ 30\%, the resulting uncertainty is not large enough to change the results at the order-of-magnitude.

As shown in Figure \ref{fig:fig5}, The spatial distributions reveal that both the CNM and LNM exhibit intricate and filamentary substructures, with the CNM showing particularly fine-scale morphology. For instance, around $(l, b)\sim(94.5^\circ, -36.5^\circ)$ and $(92.9^\circ, -36.4^\circ)$, small clumpy CNM features are seen to be surrounded by more diffuse LNM structures. Such clumpy CNM structures are found at multiple locations across the region. Typically, CNM structures have characteristic sizes of $\lesssim 1$ pc, while LNM structures tend to be comparable or somewhat larger in scale. The tendency for LNM to envelop CNM features is consistent with the all-sky phase separation reported by \citet{KalberlaHaud2018}, as well as with numerical simulations showing that LNM preferentially forms around CNM regions and along sheet-like structures surrounding dense condensations (e.g., \cite{GazolKim2010}; \cite{Saury2014}). In contrast, the WNM exhibits a more diffuse morphology with a broader and smoother spatial distribution compared to the other two phases.

Assuming optically thin conditions, the total H\,\textsc{i} mass of HLCG 92$-$35 is estimated from equation~(3) to be $\sim2200~M_\odot$, of which the CNM, LNM, and WNM contribute ~ $946~M_\odot$, $1034~M_\odot$, and $220~M_\odot$, respectively. The resulting mass fractions are CNM : LNM : WNM = 43 : 47 : 10.


\section{Discussion}

\subsection{Spatial correlation among H\,\textsc{i} phases}

Figure \ref{fig:fig5} presents the two-dimensional frequency distributions of the mass fractions of each H\,\textsc{i} phase. The mass fraction of the CNM is defined as $f_\mathrm{CNM} = N_\mathrm{CNM} / N_\mathrm{HI}$, and those of the LNM and WNM are defined analogously. The diagonal distribution seen in Figure \ref{fig:fig5}(a) indicates a strong anti-correlation between the CNM and LNM, suggesting that CNM-dominated and LNM-dominated regions are spatially complementary. 
In contrast, Figure \ref{fig:fig5}(b) shows that the distributions of the WNM and CNM are more diffuse and exhibit only a weak positive correlation. 


\citet{KalberlaHaud2018} performed a similar two-dimensional frequency analysis using the all-sky HI4PI survey and found that, on average, the LNM dominates globally with $f_\mathrm{CNM} \lesssim 0.15$, $f_\mathrm{LNM} \sim 0.55$, and $f_\mathrm{WNM} \sim 0.30$. However, in the local small-scale, filamentary components they identified, high $f_\mathrm{CNM}$ values and a strong CNM–LNM anti-correlation were observed, which are qualitatively consistent with Figure \ref{fig:fig5}. A similarly weak CNM–WNM correlation was also reported in their study.

\subsection{H\,\textsc{i} phase fraction}
The total mass ratio among the phases is CNM:LNM:WNM = 43:47:10, indicating that most of the H\,\textsc{i} in the HLCG 92$-$35 region resides in the LNM and CNM.

Previous studies based on H\,\textsc{i} absorption line observations (e.g., \cite{HeilesTroland2003b}; \cite{Roy2013b}; \cite{Murray2015}, \yearcite{Murray2018}) have reported that, near the Galactic plane, $\sim30\%$ of the H\,\textsc{i} is in the CNM and $\sim20\%$ in the LNM. In contrast, more localized observations have revealed substantial regional variations in the H\,\textsc{i} phase fractions. For example, in the North Ecliptic Pole region, CNM $\approx8\%$, LNM $\approx28\%$, and WNM $\approx64\%$ (\cite{MarchalMiville-Deschenes2021}), whereas in the North Celestial Pole Loop, \citet{Taank2022} found CNM $\approx50\%$, LNM $\approx21\%$, and WNM $\approx27\%$ for the low-velocity component associated with the loop structure. Compared with these studies, the H\,\textsc{i} phase fractions derived here are characterized by a relatively large LNM fraction, accounting for nearly half of the total H\,\textsc{i} mass, while the WNM contributes only about 10\%. The low WNM fraction suggests that the conversion of atomic gas toward colder phases is already well advanced. At the same time, the large amount of LNM indicates that a significant fraction of the gas remains outside a simple thermal equilibrium state.

Such a large LNM fraction is qualitatively consistent with theoretical and numerical studies (\cite{AuditHennebelle2005}; \cite{HennebelleAudit2007}; \cite{KimOstriker2018}), which show that external energy injection can continuously drive H\,\textsc{i} gas away from thermal equilibrium and increase the amount of gas residing in the thermally unstable regime. Numerical simulations further predict that dynamical perturbations can produce thermally unstable gas fractions of $\sim40$--$50\%$ of the total H\,\textsc{i} mass (\cite{Kobayashi2020}; \cite{Seifried2020a}, \yearcite{Seifried2020b}; \cite{Hu2025}). Although the LNM identified observationally does not necessarily correspond entirely to thermally unstable gas, the high LNM fraction found here is consistent with a scenario in which dynamical perturbations play an important role in producing and maintaining non-equilibrium H\,\textsc{i} structures. Since gas in the thermally unstable regime is expected to evolve toward the CNM on timescales of order $\lesssim1$ Myr (e.g., \cite{InoueInutsuka2012}), the observed abundance of LNM suggests that the transition from thermally unstable H\,\textsc{i} to CNM may still be ongoing.

The MBM 53, 54, and 55 complex, which encompasses the HLCG 92$-$35 region, forms part of an expanding H\,\textsc{i} shell that remains detectable today \citep{Gir1994}. The short cooling timescale of thermally unstable H\,\textsc{i} implies that a large LNM fraction would be difficult to maintain if the gas were evolving passively after a single compression event. However, the shell is still expanding at $\sim18$ km s$^{-1}$, suggesting that gas compression and turbulence driving may remain active. Under such conditions, thermally unstable gas can be continuously replenished by ongoing dynamical perturbations, helping to sustain a large LNM fraction.

This picture is also broadly consistent with the molecular-cloud formation scenario proposed by \citet{Inutsuka2015}, in which molecular clouds are not formed through a single compression event but rather through the cumulative effects of multiple compression episodes associated with expanding shells and bubbles in the interstellar medium. In this framework, repeated compressions continuously supply and restructure atomic gas, allowing cold structures to gradually grow over timescales of order $\sim10$ Myr, much longer than the cooling time of individual thermally unstable gas parcels. The ongoing expansion of the H\,\textsc{i} shell surrounding HLCG 92$-$35 therefore suggests that H\,\textsc{i} gas may still be supplied to the cloud-forming region, providing a mechanism to replenish the thermally unstable reservoir and promote the continued growth of CNM structures.

Although the replenishment rate cannot be constrained observationally, the shell kinematics indicate that dynamical perturbations may persist on timescales comparable to the cooling time and contribute to maintaining the observed phase structure. 
In this interpretation, the large LNM fraction observed toward HLCG 92$-$35 may represent an actively evolving atomic-to-molecular transition, rather than the consequence of a single past compression event. 
On the other hand, the lack of associated radio continuum and X-ray emission prevents a reliable estimate of the age of the putative supernova event and hence the duration of energy injection. Consequently, a quantitative assessment of whether the available kinetic energy is sufficient to sustain the observed LNM fraction remains beyond the scope of the present data.

\subsection{Physical properties of CNM and CO clumps}

\subsubsection{Identifying clumps: Astrodendro}

To identify clumpy structures, we employed the Python package Astrodendro, which implements the dendrogram-based analysis method described by \citet{Rosolowsky2008}. This approach was adopted because CNM structures associated with the H\,\textsc{i} phase transition exhibit inherently multiscale morphology. In our analysis, Astrodendro was applied to the integrated intensity maps of each CNM component to identify the leaf structures. The use of integrated intensity maps enables the detection of structures characterized by high spectral intensity. Because Astrodendro identifies hierarchical structures directly from the data, and because most of the CNM integrated intensities satisfy the $5~\sigma$ detection threshold adopted in this study (see below), the impact of parameter selection on the results is expected to be limited (\cite{Goodman2009}; \cite{Marchal2021}, and see also Appendix~\ref{appsec:dendrogram}).

\subsubsection{Derivation of Physical Properties}

To prevent the extraction of spurious structures caused by noise during clump identification, the following parameters were adopted in Astrodendro.

First, the minimum intensity required for a structure to be identified was set to \verb|min_value| = 3.56 [$\mathrm{K~km~s^{-1}}$], corresponding to the $5~\sigma$ detection limit in the integrated intensity map of the GALFA–HI dataset. Structures below this threshold were excluded from the analysis. In addition, to ensure that a structure is identified as an independent leaf separated from neighboring structures, its peak intensity was required to exceed the merge level with adjacent structures by at least \verb|min_delta| = 3.56 [$\mathrm{K~km~s^{-1}}$]. Furthermore, to guarantee that each identified leaf has sufficient spatial extent, we adopted \verb|min_npix| = 25. Given the $1'$ sampling interval and the effective beam size of $\sim 4'$ in the GALFA-HI data, 25 pixels corresponds to an area approximately twice the beam size. This criterion prevents small-scale fluctuations significantly smaller than the beam size from being mistakenly identified as leaves.

For each identified clump, we derived its size (diameter), mass, and mean density.
The clump size was defined from the area A occupied by the identified leaf as the diameter of a circle with the same area, $L = 2\sqrt{{A}/{\pi}}$ .
The clump mass $M_\mathrm{CNM}$ was calculated as

\begin{equation}
    M_\mathrm{CNM}=m_\mathrm{H}\sum [d^2\Omega N_\mathrm{CNM}]
\end{equation}
where $m_\mathrm{H}$ is the mass of a hydrogen atom, $d$ is the distance to the cloud, $\Omega$ is the solid angle per pixel ($1~\mathrm{arcmin}^2$), and $N_\mathrm{CNM}$ is the column density of the CNM within the region occupied by the leaf structure. The $\Sigma$ represents the sum over all pixels comprising the clump. Assuming the gas is optically thin, $N_\mathrm{CNM}$ was derived using Equation (3).

The mean number density of each clump was then estimated from its mass $M_\mathrm{CNM}$ and diameter $L$. As before, this estimate should be regarded as a lower limit.

\begin{equation}
    n_\mathrm{CNM}=\frac{M_\mathrm{CNM}}{m_\mathrm{H}\,(4/3)\pi(L/2)^3}\quad [\mathrm{cm^{-3}}],
\end{equation}
A dendrogram analysis was also performed on the $^{12}$CO ($J$=1--0) integrated intensity map of the HLCG 92$-$35 region using Astrodendro. The adopted parameters were \verb|min_value| = 0.9 [$\mathrm{K~km~s^{-1}}$] ($\approx 3\sigma$), \verb|min_delta| = 0.6 [$\mathrm{K~km~s^{-1}}$] ($\approx 2\sigma$), and \verb|min_npix| = 2. Structures with at least 2 pixels were identified as clumps, and this definition is based on the previous study of CO clump analysis in the HLCG 92$-$35 region by \citet{Yamamoto2003}.
The clump mass was calculated as

\begin{equation}
    M_\mathrm{^{12}CO}=\mu m_\mathrm{H}\int [d^2\Omega N_\mathrm{H_2}]dS
\end{equation}

where $\Omega$ is the solid angle per pixel, corresponding to $4~\mathrm{arcmin}^2$. The molecular hydrogen column density, $N_\mathrm{H_2}$, was derived from the integrated intensity of $^{12}$CO using a CO-to-H$_2$ conversion factor of $1.3\times 10^{20}~\mathrm{cm^{-2}/(K~km~s^{-1})}$ as measured for high-latitude molecular clouds by \citet{CottenMagnani2013}. The mean molecular weight per hydrogen molecule, $\mu$, was assumed to be 2.0, corresponding to a pure H$_2$ composition.

\subsubsection{Physical properties of CNM clumps}

A total of 1792 CNM clumps were identified. Many of these clumps overlap along the line of sight and therefore cannot be clearly separated in the integrated intensity maps shown in Figures \ref{fig:fig1} and \ref{fig:fig4}.

Figure \ref{fig:fig6}(a) shows the diameter distribution of the CNM clumps, which ranges from $1.0$ to $9.3$ pc, with an average diameter of $1.9 ~\pm ~0.9$ pc.


The data used in this study achieve a higher spatial resolution than previous studies (e.g., \cite{Marchal2021}), enabling the analysis of CNM structures on smaller spatial scales. \citep{Marchal2021} reported an average CNM structure size of 6.4 pc based on observations with a spatial resolution of $\sim$ 3.2 pc. In contrast, this study identified CNM structures with smaller sizes. This may be because the improved spatial resolution allowed us to resolve CNM structures that had previously been recognized as single entities into multiple sub-pc-scale components.
Nevertheless, even with the spatial resolution of this study ($\sim 0.2$ pc), the observed CNM structures may still be affected by resolution limitations. For example, numerical simulations by \citet{Kobayashi2023} showed that the size distribution of CNM clumps ranges from 0.03 pc to several pc, with a peak around 0.1 pc. This suggests that observations with even higher spatial resolution may reveal smaller CNM structures. Therefore, further high-resolution observations are required to better understand the internal structure of the CNM.

Figure \ref{fig:fig6}(b) shows the number density distribution of the CNM clumps. The derived densities range from 4~$\mathrm{cm^{-3}}$ to $460~\mathrm{cm^{-3}}$, with an average value of $81~\pm ~67 ~\mathrm{cm^{-3}}$. This result is in good agreement with the mean CNM number density of $76~\mathrm{cm^{-3}}$ estimated by \citet{Kalberla2025} for high galactic  latitude regions based on absorption line observations. This indicates that our results are consistent with those derived from absorption data. High-density CNM clumps are considered potential sites for the onset of molecular cloud formation. As shown in Figure \ref{fig:fig6}(b), clumps with densities below $100~\mathrm{cm^{-3}}$ constitute the majority, while only 118 clumps (7\% of the total) have densities above $200~\mathrm{cm^{-3}}$. As the CNM becomes denser, molecular hydrogen is expected to form within the clumps, suggesting that the CNM density tends to saturate at several hundred $\mathrm{cm^{-3}}$.  Indeed, numerical simulations of H$_2$ formation by \citet{Valdivia2016} demonstrate that most H$_2$ molecules form at densities of a few hundred $\mathrm{cm^{-3}}$. Consequently, the CNM density is expected to reach a natural plateau as the gas transitions into the molecular phase. Other theoretical simulation studies  (e.g., \cite{Fukui2018}) have also shown that the maximum CNM density reaches  $\sim 10^3~\mathrm{cm^{-3}}$. The spatial resolution of the simulations in \citet{Fukui2018} was $0.02$ pc, whereas the observational data used in this study are affected by beam dilution. Consequently, the maximum CNM densities derived here are expected to be slightly lower than the simulation values, however generally fall within the upper end of the $10^2~\mathrm{cm^{-3}}$ range.

\subsubsection{Mass spectrum of CNM clump and CO clump}

The derived CNM clump mass spectrum is shown in Figure \ref{fig:fig7}(a). The masses of the CNM clumps range from 4.0 $\times10^{-3}~M_\odot$ to 1.7 $~M_\odot$. The detection limit of $N_\mathrm{H\,I}$ in the GALFA–HI DR2 data is $\approx 6.5\times10^{18}~\mathrm{cm^{-2}}$ \citep{Peek2018}, which corresponds to a mass detection threshold of $\approx 2.47\times10^{-3}~M_\odot$ derived using the Astrodendro. The mass distribution of the clumps, expressed as the number per unit mass $dN/dM \propto M^{-\gamma}$, follows a power-law form with $\gamma = 1.76 \pm 0.11$ for clump masses in the range of $0.05 < M < 0.5~M_\odot$ and $\gamma = 3.11 \pm 0.52$ for $M > 0.5~M_\odot$. The high-mass end of the distribution is subject to large uncertainties owing to the small number of detected clumps.

The spectral slope obtained for the intermediate-mass range is in good agreement with values reported in simulations of WNM inflows (e.g, \cite{Heitsch2005}; \cite{HennebelleAudit2007}; \cite{InoueInutsuka2012}; \cite{Kobayashi2023}). \citet{HennebelleAudit2007} investigated the cooling of turbulent H\,\textsc{i} gas flows using two-dimensional simulations and demonstrated that CNM clumps formed via thermal instability exhibit hierarchical and statistically self-similar structures. Using an analytical approach based on Press–Schechter theory, they showed that if the spectrum of initial density fluctuation has an exponent of $n = 11/3$ (Kolmogorov type), the resulting mass distribution of clumps formed through thermal instability follows $\gamma = 2 - (n-3)/3 \simeq 1.8$. The close agreement between the spectral index derived in this study and this theoretical prediction is consistent with the interpretation that the intermediate-mass CNM clumps were formed during the growth of density structures driven by thermal instability.

Figure \ref{fig:fig7}(b) shows the mass distribution of CO clumps in the HLCG 92$-$35 region identified in Section 5.3.2. 
The masses of the CO clumps range from $0.16$ to $24~M_\odot$. The $3\sigma$ detection limit for the H$_2$ column density is $1.2\times10^{20}~\mathrm{cm^{-2}}$, corresponding to a mass detection limit of $\sim0.12~M_\odot$ for the adopted clump-identification criterion ($\mathrm{min_npix}=2$; red dashed line in the figure). The CO clump mass distribution follows a single power law, $dN/dM \propto M^{-1.75\pm0.08}$, for $M > 0.4~M_\odot$. This slope is comparable to that found in the intermediate-mass range of CNM clumps. This similarity suggests that molecular clouds may form while inheriting the mass distribution of CNM structures.


The steepening of the mass distribution in the range $M \gtrsim 0.5~M_\odot$ may reflect the evolutionary stage of massive CNM clumps transitioning into molecular clouds. As these massive CNM clumps grow, their internal densities increase, promoting the formation of H$_2$. As H$_2$ formation proceeds, the fraction of mass contained in atomic hydrogen decreases over time, leading to a sharp decline in the number of massive clumps remaining within the CNM regime. Consequently, the steepening of the mass spectrum can be interpreted as a statistical signature of the ongoing transition from CNM to H$_2$. Indeed, massive clumps are generally denser, and the rate coefficient for H$_2$ formation on dust grain surfaces is known to increase rapidly with the density of atomic hydrogen \citep{TielensHollenbach1985}. Therefore, massive and dense CNM clumps are expected to evolve into molecular clouds on relatively short timescales, resulting in a decrease in the mass observed as CNM. Supporting this interpretation, \citet{Valdivia2016} analyzed simulations of molecular cloud formation in which clumps were identified based on the total hydrogen number density ($n=n_\mathrm{CNM}+2n_\mathrm{H_2}$) and the molecular hydrogen fraction, $f(\mathrm{H}_2)=2n_\mathrm{H_2}/n_\mathrm{CNM}$, within each clump was statistically evaluated. They found that clumps with masses of $\sim0.1~M_\odot$ already exhibit $f(\mathrm{H}_2) > 0.2$, while more massive clumps with $M \gtrsim 1~M_\odot$ typically reach $f(\mathrm{H}_2) \gtrsim 0.8$. This  demonstrates a systematic increase of the molecular fraction with increasing clump mass. Note, however, that the clumps in their study were defined using a relatively high density threshold of $n=1000~\mathrm{cm^{-3}}$. Even for clumps with $f(\mathrm{H}_2) \gtrsim 0.8$, this corresponds to $n_\mathrm{CNM}\gtrsim 200~\mathrm{cm^{-3}}$, and the analysis therefore focuses on comparatively dense clumps.

Another possible explanation is that the massive CNM clumps continue to grow through mergers with neighboring CNM structures and accretion of surrounding atomic gas. In this case, mass is progressively transferred from the ordinary clump population into a small number of exceptionally massive cloud complexes, leading to a deficit of clumps relative to the extrapolated power law at the high-mass end. Numerical simulations (e.g., \cite{InoueInutsuka2012}; \cite{Kobayashi2023}) have shown that CNM clumps can merge to form larger cloud complexes. This suggests that clump merging may contribute to the evolution of the high-mass end of the mass spectrum. At present, however, the data analyzed in this study do not allow us to distinguish whether the steepening is primarily caused by H$_2$ formation, clump growth through mergers, or a combination of both processes.

On the other hand, the Astrodendro analysis itself may contribute to an underestimation of the number of massive clumps. Specifically, the choice of the parameter \verb|min_delta| could cause small noise-induced substructures to be identified as independent clumps, thereby artificially lowering the count of massive ones. However, the overall trend in the mass distribution remains consistent even when \verb|min_delta| is varied, suggesting that the influence of such analysis biases is likely limited (see Appendix~\ref{appsec:dendrogram}).

The fact that intermediate-mass CNM clumps and CO clumps exhibit similar power-law distributions may suggest that hierarchical density structures generated by thermal instability and turbulent compression are, at least partially, inherited from the CNM phase to the CO phase. The observed self-similarity in the mass distribution could indicate that such hierarchical structures are established during the formation of CNM from the WNM and subsequently retained during molecular cloud formation.

Deviations from the power-law at the low-mass end ($M \lesssim 0.05~M_\odot$) are likely caused by the limited spatial resolution of the observations. In three-dimensional simulations by \citet{InoueInutsuka2012} and \citet{Kobayashi2023}, the mass function continues to follow a power-law even below $0.05~M_\odot$. In contrast, the present study shows clear deviations in this low-mass regime, indicating a relative deficiency of low-mass CNM clumps (i.e., smaller-sized structures) compared with the simulations. This discrepancy suggests that low-mass clumps may be missed observationally or that multiple unresolved clumps are blended into a single detected structure. \citet{AuditHennebelle2010} examined clump masses at various spatial resolutions and found that the lower-mass turnover from the power law shifts to smaller masses as resolution improves. The spatial resolution of their simulations ($\sim0.02~\mathrm{pc}$) is about an order of magnitude finer than that of the GALFA–H I data used in this study ($\sim0.2~\mathrm{pc}$). Numerical simulations predict that numerous low-mass CNM clumps exist on sub-0.1 pc scales, which fall below the resolution limit of the current observations. Consequently, these small clumps are likely unresolved or undetected in our data. Therefore, the apparent deficiency of low-mass clumps arises primarily from observational limitations and is expected to be mitigated by future H\,\textsc{i} observations with higher angular resolution.

\subsection{Velocity dispersion of the CNM spectrum}
\label{subsec:velocity dispersion}

The line width of the separated CNM components is generally broader than the thermal line width expected from the canonical CNM temperature. The spectral velocity dispersion, $\sigma$, is related to the gas kinetic temperature, $T_\mathrm{k}$, by

\begin{equation}
T_\mathrm{k}=
\frac{m_\mathrm{H}}{k_\mathrm{B}}\sigma^2=121\,\sigma^2
\end{equation}

For reference, the thermal velocity dispersion of H\,\textsc{i} gas at 100 K is $\sim 0.91~\mathrm{km~s^{-1}}$.

The observed velocity dispersion includes both thermal and nonthermal contributions. The total dispersion, $\sigma_\mathrm{obs}$, can be expressed as
\begin{equation}
\sigma_\mathrm{obs}^2=\sigma_\mathrm{th}^2+\sigma_\mathrm{nt}^2=\frac{k_\mathrm{B}T_\mathrm{k}}{m_\mathrm{H}}+\sigma_\mathrm{nt}^2
\end{equation}
where $\sigma_\mathrm{th}$ and $\sigma_\mathrm{nt}$ denote the thermal and nonthermal velocity dispersions, respectively.

Extensive H\,\textsc{i} absorption-line observations have enabled direct measurements of the CNM spin temperature.
For example, the BIGHICAT catalog, which compiles Galactic H\,\textsc{i} absorption-line data, contains 1370 detected CNM components and shows that the CNM spin temperature distribution peaks in the range of 50–200 K \citep{McClure-Griffiths2023}. Only a small fraction of components exhibit higher temperatures: 5.0\% exceed 500 K, and 1.7\% exceed 1000 K. Owing to the relatively high density of the CNM, frequent collisions with electrons, ions, and hydrogen atoms drive the hyperfine level populations of the 21 cm line toward thermal equilibrium. Consequently, the spin temperature, $T_\mathrm{s}$, is expected to be approximately equal to the kinetic temperature, $T_\mathrm{k}$, allowing a direct comparison between the two.


Figure \ref{fig:fig8} shows the distribution of line widths, expressed as equivalent temperatures, for all CNM components decomposed by ROHSA. The upper axis converts line width to temperature. The temperatures inferred from CNM line widths range from 0 to 2500 K, with most components concentrated below 1200 K. A prominent peak appears around 300 K, and the mean value is $\sim$ 530 K. This is significantly higher than the typical CNM temperatures suggested by simulations or absorption-line measurements, indicating that the line-width-derived temperatures are not purely thermal in origin but include a substantial contribution from nonthermal velocity dispersion, consistent with the absorption-line results.

Several studies have investigated the velocity dispersion of CNM components. For example, \citet{Marchal2021} derived CNM temperatures from Gaussian line widths using an approach similar to that adopted in this work. They reported a mean velocity dispersion of $1.6 \pm 1.1~\mathrm{km~s^{-1}}$, which is comparable to the spectral resolution of their data ($1.32~\mathrm{km~s^{-1}}$). This suggests that the narrowest CNM components may not have been fully resolved and that the derived line widths could be influenced by the finite spectral resolution.

The GALFA-HI data used in the present study provide a substantially higher spectral resolution of $0.184~\mathrm{km~s^{-1}}$, sufficient to resolve the thermal line width expected for CNM gas with temperatures below $\sim100$ K. Therefore, uncertainties associated with instrumental spectral resolution are expected to be reduced relative to previous studies.

Nevertheless, higher spectral resolution does not eliminate all sources of uncertainty. The finite angular resolution of the observations implies that a single telescope beam may contain multiple CNM structures. In such cases, the observed spectrum represents the superposition of several unresolved velocity components rather than the spectrum of an individual CNM structure. Because the H\,\textsc{i} 21 cm line is intrinsically broader than molecular lines, unresolved components with similar radial velocities may overlap in velocity space. Such blending can artificially broaden the fitted components and contribute to the observed velocity dispersion. 

Consequently, part of the excess line width reported here may arise from these observational effects rather than solely from the intrinsic properties of the gas. The potential impact of finite angular resolution is discussed further in Section~5.6.

\subsection{Central velocity offset between CO and CNM}
\label{subsec:velocity offset}
If the CNM serves as a building block of molecular clouds, the velocities of CNM and CO gas are expected to be closely related. We therefore examine the central velocity offsets between the $^{12}$CO($J=1$--0) emission line and the associated CNM components.

Because multiple CNM components can exist along a given line of sight, the velocity offset was computed by selecting the CNM component whose central velocity is closest to that of each CO emission line. Only pixels with CO peak intensities above $0.2$ K were included in the analysis. For CO spectra with multiple peaks, only the strongest component was considered. Each spectrum was fitted with Gaussian functions, and poorly fitted cases were excluded.
Figure \ref{fig:fig9} shows representative examples of the CNM--CO velocity offset measurements. The blue spectra correspond to CNM components (dashed lines indicate all fitted Gaussian components), while the gray spectra correspond to CO emission. 

Figure \ref{fig:fig10} presents the cumulative distribution of the central velocity difference between the $^{12}$CO($J=1$--0) emission line and the CNM components. The distribution shows that CNM and CO are not always kinematically coincident. Offsets exceeding $0.5~\mathrm{km~s^{-1}}$ are found in approximately $60\%$ of the sample, while offsets greater than $1.0~\mathrm{km~s^{-1}}$ occur in approximately $10\%$.

It should be emphasized that this pairing procedure does not guarantee a unique physical correspondence between individual CNM and CO components along the line of sight. Therefore, this analysis is not intended to identify one-to-one dynamical associations, but rather to characterize the statistical distribution of velocity offsets between the CNM and CO phases.

\citet{Park2023} also reported velocity offsets between CO and CNM along several lines of sight based on H\,\textsc{i} absorption observations toward 58 positions across the sky. Their analysis revealed velocity differences ranging from $0.01$ to $4.3~\mathrm{km~s^{-1}}$, with a median of $0.4~\mathrm{km~s^{-1}}$. Furthermore, a velocity difference of $0.06$--$2.64~\mathrm{km~s^{-1}}$ (median $0.5~\mathrm{km~s^{-1}}$) was also found between CNM and OH absorption components. 

In contrast to absorption-line studies that probe individual lines of sight, the present analysis provides spatially resolved information on the distribution of velocity offsets. However, no clear large-scale spatial coherence or systematic gradient is identified within the current dataset. This suggests that the observed offsets are not dominated by global kinematic patterns such as Galactic rotation or motions associated with the MBM53–55 complex. Instead, they are more likely related to unresolved sub-pc scale structure in the CNM, reflecting multiple kinematically distinct components.

These results suggest that molecular clouds trace only a subset of CNM structures. The CNM observed in emission likely consists of multiple kinematically distinct subcomponents, and only a fraction of them satisfy the physical conditions required for CO formation and therefore contribute to CO-bearing gas.


\subsection{Interpretation of the CNM line broadening}

The CNM line broadening discussed in Section \ref{subsec:velocity dispersion}, together with the velocity offsets between CO and CNM presented in Section \ref{subsec:velocity offset}, suggests that the CNM may contain unresolved substructure. In this section, we discuss the possibility that the nonthermal contribution to the CNM linewidth originates from very small-scale internal structures.

At the angular resolution of GALFA-HI, as illustrated in Figure \ref{fig:fig11}(a), molecular clouds traced by the $^{12}$CO($J=1$--0) line occupy only a limited portion of the velocity range associated with a single CNM component. If the observed velocity distribution directly reflects the physical distribution of gas along the line of sight, this would imply that molecular clouds form preferentially within only part of an extended CNM structure. However, an alternative interpretation is possible. As schematically shown in Figure \ref{fig:fig11}(b), the observed CNM component may actually consist of several small CNM cloudlets with slightly different velocities that are blended together within a single telescope beam. Molecular clouds may then form only in a subset of these cloudlets.
The velocity resolution of GALFA-HI ($0.18~{\rm km\,s^{-1}}$) is sufficient to resolve the narrow spectral features expected from CNM gas. In contrast, its spatial resolution ($4'$; $\sim0.2$ pc at 150 pc) is substantially larger than the smallest CNM structures predicted by numerical simulations. For example, \cite{Kobayashi2023} reported CNM clumps as small as $\sim0.03$ pc in MHD simulations. Consequently, multiple CNM cloudlets may be blended within a single telescope beam. Although each cloudlet may possess a narrow intrinsic linewidth, velocity differences among the cloudlets can be merged into a broader spectral profile, producing an apparent nonthermal linewidth. In this sense, insufficient spatial resolution can effectively degrade the velocity resolution of the observations, because the measured spectrum may represent the superposition of multiple unresolved velocity components within the telescope beam rather than the spectrum of an individual CNM structure.

In addition to beam blending, the superposition of structures along the line of sight may also contribute to the observed CNM linewidths. Since H\,\textsc{i} is widely distributed in the local interstellar medium, multiple CNM structures at different distances may overlap in velocity space when their radial velocities are similar. Such superposition can broaden the observed spectral profiles and complicate the interpretation of individual Gaussian components. The BIGHICAT absorption-line data provide an additional constraint on this issue. Among the 1370 CNM components identified in the BIGHICAT catalog, 947 components ($\sim$69\%) have spin temperatures of $T_{\rm s}<250$ K. The median FWHM of these absorption components is $3.1~\mathrm{km\,s^{-1}}$, and approximately 73\% have linewidths narrower than $5~\mathrm{km\,s^{-1}}$. Because absorption measurements use background continuum sources as sampling beams, they effectively probe the CNM at much higher angular resolution than H\,\textsc{i} emission observations. While such measurements do not eliminate confusion along the line of sight, they largely avoid spatial averaging within the telescope beam. Although line-of-sight blending may still contribute to the observed linewidths, the narrower linewidths found in absorption studies suggest that, at least on the spatial scale of GALFA-HI, beam blending of unresolved CNM structures within the telescope beam may be a major contributor to the linewidth broadening observed in emission.

Under the cloudlet interpretation described above, the velocity offsets observed between the CO and CNM peaks do not necessarily imply a physical displacement between molecular and atomic gas. Instead, they may arise because CO emission traces only a subset of the unresolved CNM cloudlets contributing to the broader CNM profile. It should be noted, however, that the CO structures analyzed in this study may also be spatially smoothed by the telescope beam, implying that even smaller molecular substructures could be embedded within the CNM and remain unresolved.

From Figure \ref{fig:fig8}, the CNM line width is typically several $\mathrm{km~s^{-1}}$. In comparison, the sound speed in CNM is $c_s=\sqrt{k_\mathrm{B}T/\mu m_\mathrm{H}}\sim 1~ \mathrm{km~s^{-1}}$ for $T \sim 100$ K. 
This implies that many of the observed CNM components contain non-thermal motions exceeding the CNM sound speed, with non-thermal linewidths often larger than $1~\mathrm{km~s^{-1}}$.
If this non-thermal broadening is interpreted as internal bulk motions within a single coherent CNM cloud, such supersonic motions would rapidly dissipate through shock formation, making it difficult to sustain them over long timescales. An alternative interpretation is that the observed line profile results from the superposition of multiple small CNM clumps unresolved within the telescope beam, both along the line of sight and across the beam. In this case, these clumps are embedded in a warmer phase such as the WNM or LNM, where the sound speed is $\sim 10~\mathrm{km~s^{-1}}$. The relative motions between CNM clumps would then be subsonic with respect to the surrounding WNM and LNM, so that shock dissipation is not efficiently triggered in the ambient gas. As a result, the observed velocity dispersion can persist for extended periods without rapid decay.

This interpretation is supported by theoretical studies (e.g., \cite{KoyamaInutsuka2002}; \cite{AuditHennebelle2005}; \cite{Heitsch2005}; \cite{Kobayashi2020}, \yearcite{Kobayashi2022}), which demonstrate that the CNM develops sub pc scale filamentary or clumpy structures through thermal instability. These simulations indicate that internal velocity dispersions within individual CNM clumps remain subsonic ($< 1~\mathrm{km~s^{-1}}$), while relative velocities between clumps typically reach several~$\mathrm{km~s^{-1}}$.

If molecular clouds are formed from such small CNM clumps, their motions are expected to be inherited by the resulting molecular clouds. Observational studies have indeed reported small scale structures in the interstellar medium. \citet{Sakamoto2002} and \citet{SakamotoSunada2003} identified $\sim$ 10000 au scale substructures at the edges of high latitude clouds (MBM 54 and MBM 55), and in the Heiles Cloud 2 region in Taurus. Furthermore, \citet{Tachihara2012}, based on $^{12}$CO ($J=1$--0 and 3--2) observations of the LDN 204 molecular cloud interacting with an H\,\textsc{ii} region, detected small cloudlets with sizes of $10^{3}$--$10^{4}$ au. These cloudlets exhibit low internal velocity dispersions but relative motions of several $\mathrm{km~s^{-1}}$ between individual structures, which suggests that such cloudlets may contribute to the non-thermal linewidths observed in molecular clouds. However, it remains unclear whether these structures originated from CNM clumps. To test this scenario, future high-resolution combined H\,\textsc{i} and CO observations will be essential. In particular, observations of low-column-density regions such as cloud edges, where line-of-sight confusion is expected to be reduced, may provide the most direct clues to the physical relationship between CNM cloudlets, molecular cloudlets, and the origin of the observed velocity structure.

\section{Summary}

Using ROHSA, we decomposed the GALFA-HI data of the molecular cloud formation region HLCG 92$-$35, observed with the Arecibo Observatory, into three components. These components were classified as Cold Neutral Medium (CNM), Lukewarm Neutral Medium (LNM), and Warm Neutral Medium (WNM) in order of decreasing line width.
Nine Gaussian components were employed for the spectral decomposition, consisting of two for the WNM, one for the LNM, and six for the CNM. Furthermore, Astrodendro was applied to identify clump-like structures within the CNM and to derive their physical parameters. For comparison with the CNM, we used $^{12}$CO ($J$=1--0) data obtained with the NANTEN telescope.

These analyses reveal that the observed CNM structures are broadly consistent with the picture suggested by previous theoretical studies. In particular, the indications of sub pc scale structure emphasize the importance of future high-resolution H\,\textsc{i}/CO observations.

Our main findings are summarized as follows:

\begin{enumerate}

  \item In the HLCG 92$-$35 region, the H\,\textsc{i} mass ratio among the three phases was found to be CNM:LNM:WNM = 43:47:10. While it remains uncertain whether all of the LNM is thermally unstable, a substantial fraction of the gas is found to reside in the intermediate-temperature phase. This result suggests that the gas is driven out of thermal equilibrium by turbulence, and that the thermodynamic state of H \textsc{i} may still be influenced by residual energy input from past shock-driven processes. This interpretation is consistent with previous observational and theoretical studies reporting a significant fraction of gas in the intermediate phase.
  
  \item A total of 1792 CNM clumps were identified by the Astrodendro, with an average size of $1.9$ pc and an average density of $\sim$ 81 $\mathrm{cm^{-3}}$. These small-scale ($\sim$ 1 pc), high-density structures are consistent with predictions from theoretical simulations. While most CNM clumps are relatively diffuse, a subset of high-density clumps likely represents the earliest stages of molecular cloud formation.

  \item The mass spectrum of CNM clumps follows a power law of $dN/dM \propto M^{-1.76 \pm 0.11}$ for $0.05 < M < 0.5~M_\odot$, but steepens significantly at higher masses. CO clumps ($M > 0.4~M_\odot$) exhibit a power-law slope of $M^{-1.75 \pm 0.08}$, which is consistent with that of the intermediate-mass CNM. This similarity in the mass distributions suggest that molecular clouds inherit the hierarchical structure of the CNM. This result implies that the self-similar nature of the gas is maintained throughout the evolutionary transition from atomic to molecular phases, likely governed by thermal instability and turbulent compression.

  \item Analysis of CNM linewidths revealed a substantial non-thermal component. The velocity offsets between CNM and CO (typically $0.5$-$1~\mathrm{km~s^{-1}}$) appear to be local phenomena arising from structural inhomogeneities within the CNM. These features imply that the CNM may consist of unresolved, subsonic small scale clouds moving collectively as aggregates potentially representing the earliest phase of molecular cloud formation.
\end{enumerate}

\section*{Acknowledgements}
This publication utilizes data from Galactic ALFA H\,\textsc{i} (GALFA-HI) survey data set obtained with the Arecibo L-band Feed Array (ALFA) on the Arecibo 305m telescope. The Arecibo Observatory is a facility of the National Science Foundation (NSF) operated by SRI International in alliance with the Universities Space Research Association (USRA) and UMET under a cooperative agreement.  The GALFA-HI surveys are funded by the NSF through grants to Columbia University, the University of Wisconsin, and the University of California.
The NANTEN telescope was operated under a mutual agreement between
Nagoya University and the Carnegie Institution of Washington,
with additional support from Japanese public donors and companies.
This work was financially supported by JST SPRING, Grant Number JPMJSP2125.
The author Y.M. would like to thank the “THERS Make New Standards Program for the Next Generation Researchers.”

\appendix

\section{Supplementary Maps of H\,\textsc{i} Components and H\,\textsc{i}/CO Clumps}
Figure \ref{fig:fig13} shows the integrated intensity maps of the nine H\,\textsc{i} components separated using ROHSA.
The color scale differs for each component.
For the six CNM phases, Figure \ref{fig:fig14} presents the CNM clump structures identified in each component, overlaid on their corresponding parent integrated intensity maps.
Figure \ref{fig:fig15} shows the CO clumps in the HLCG 92$-$35 region identified using the Astrodendro.
Clump identification was performed on the entire $^{12}\mathrm{CO}(J=1\text{--}0)$ data set, but only clumps located within the HLCG 92$-$35 region shown in Figure \ref{fig:fig15} were used in the analysis. Clumps lying on or near the boundary of the analyzed region were excluded. In Figure \ref{fig:fig15}, CNM clump structures are shown as contours overlaid on the $^{12}\mathrm{CO}(J=1\text{--}0)$ integrated intensity map, while plus symbols mark the centroids of the selected clumps.

\section{Robustness of the Gaussian Decomposition}
\label{appsec:ROHSA}

To assess the robustness of the Gaussian decomposition, we examined the dependence of the results on the adopted ROHSA parameters. We considered models that provide comparably good fitting performance (see Figure \ref{fig:fig1b}), with $N = 8$, 9, and 10 Gaussian components and regularization parameters $\lambda_a = \lambda_\mu = \lambda_\sigma = 1$, 10, 100, and 1000, while fixing $\lambda'_\sigma = 1$. For all of these models, the Gaussian components remain separated into three major clusters in parameter space (Figure \ref{fig:fig16}, \ref{fig:fig17}, \ref{fig:fig18}). This result demonstrates that the identification of three distinct populations is largely insensitive to the specific choice of ROHSA parameters. However, the mass fraction associated with Cluster B, corresponding to the intermediate-linewidth population, can vary by $\sim$ 10\% depending on the adopted model. In some cases, the mean linewidth of Cluster B is also reduced to $\sim 13~\mathrm{km~s^{-1}}$, smaller than that obtained in the fiducial model. This behavior likely reflects the fact that Cluster B occupies an intermediate region between Clusters A and C in parameter space. As a result, some Gaussian components may be assigned to different clusters depending on the details of the decomposition, leading to increased uncertainty in the separation of Cluster B.In contrast, the mass fraction of Cluster C remains nearly unchanged among the tested models. This suggests that the observed variation primarily arises from components that are alternatively classified as Cluster A or Cluster B, rather than from any significant redistribution involving Cluster C. Therefore, although the precise properties of Cluster B show some dependence on the adopted ROHSA parameters, the existence of three major Gaussian-component populations is a robust and consistent result.

\section{Dependence of the Dendrogram Analysis on Parameter Selection}
\label{appsec:dendrogram}
To evaluate the robustness of the dendrogram analysis, we investigated the effects of varying the dendrogram parameters \verb|min_value|, \verb|min_delta|, and \verb|min_npix| on the derived clump properties and mass spectra. The parameter \verb|min_npix| determines the minimum size of structures that can be identified. Under the condition of \verb|min_value = min_delta| = $5\sigma_\mathrm{rms}$, increasing \verb|min_npix| from 16 to 25 changes the minimum detectable clump size from approximately 0.20 pc to 0.25 pc. If \verb|min_npix| is chosen too small, noise fluctuations may be incorrectly identified as physical structures. In contrast, \verb|min_value| and \verb|min_delta| primarily affect the minimum detectable clump mass. For example, with \verb|min_npix| = 25, increasing these parameters from $3\sigma_\mathrm{rms}$ to $5\sigma_\mathrm{rms}$ raises the minimum detectable mass from approximately $2\times10^{-3},M_\odot$ to $4\times10^{-3},M_\odot$.

To examine the stability of the derived mass spectrum, we fixed \verb|min_npix| = 25 and varied \verb|min_value| and \verb|min_delta| between $3\sigma_\mathrm{rms}$ and $7\sigma_\mathrm{rms}$. As shown in Figure \ref{fig:fig19}(a), the low-mass end of the mass spectrum changes due to detection limits and spatial-resolution effects. However, the slope of the mass function in the intermediate-mass regime remains nearly unchanged, with a typical value of $\sim -1.8$. These results indicate that the derived mass function is largely insensitive to the choice of dendrogram parameters, except at the low-mass end where completeness effects become important. Based on this analysis, we adopted \verb|min_npix| = 25 and \verb|min_value = min_delta| = $5\sigma_\mathrm{rms}$, which provide a reasonable balance between sensitivity and robustness.

A similar dependence on the dendrogram parameters is found for the CO clump identification. As in the CNM analysis, increasing \verb|min_value|, \verb|min_delta|, or \verb|min_npix| preferentially suppresses the low-mass end of the mass spectrum because low-mass clumps are more strongly affected by the spatial resolution and sensitivity limits. Figure \ref{fig:fig19}(b) shows the mass spectra derived with \verb|min_value| = 3$\sigma$, \verb|min_delta| = 2$\sigma$, and \verb|min_npix| = 2, 3, and 4. For \verb|min_npix| = 2 and 3, the mass functions exhibit similar slopes of $\sim -1.8$, indicating that the overall shape of the distribution is robust against moderate changes in the size threshold. In contrast, the mass spectrum obtained with \verb|min_npix| = 4 deviates significantly from the other cases. This is primarily because increasing \verb|min_npix| reduces the number of identified clumps, leading to larger statistical fluctuations in the derived distribution. The numbers of identified clumps are 170, 113, and 89 for \verb|min_npix| = 2, 3, and 4, respectively.

Based on these results, we adopted \verb|min_npix| = 2 for the CO analysis. This choice minimizes the difference in the effective size threshold relative to the H\,\textsc{i} analysis while maintaining sufficient statistics for the mass-spectrum analysis. The adopted threshold is also consistent with the clump-identification criterion used by \citet{Yamamoto2003}, in which molecular clumps were required to extend over at least two spatial pixels.

\clearpage

\begin{figure}
 \begin{center}
  \includegraphics[width=\columnwidth]{Fig2_LP_phase_with_hist_v2.png} 
 \end{center}
\caption{(Top) Distribution of 1,575,000 Gaussian components decomposed by ROHSA in the FWHM–peak intensity plane. Colors indicate the phase classification: CNM (blue), LNM (purple), and WNM (red). Contours represent the point density.
(Bottom) Histogram of the Gaussian FWHM values for each phase, with the vertical axis shown on a logarithmic scale. The FWHM distributions are not completely separated by phase and exhibit partial overlap between the components.
{Alt text: Two vertically stacked panels showing properties of Gaussian components derived from neutral hydrogen data. The top panel presents peak intensity as a function of full width at half maximum for one hundred fifty five thousand components. The bottom panel shows the distribution of full width at half maximum values as a histogram on a logarithmic count scale.}}\label{fig:fig2}
\end{figure}

\begin{figure}
 \begin{center}
  \includegraphics[width=\columnwidth]{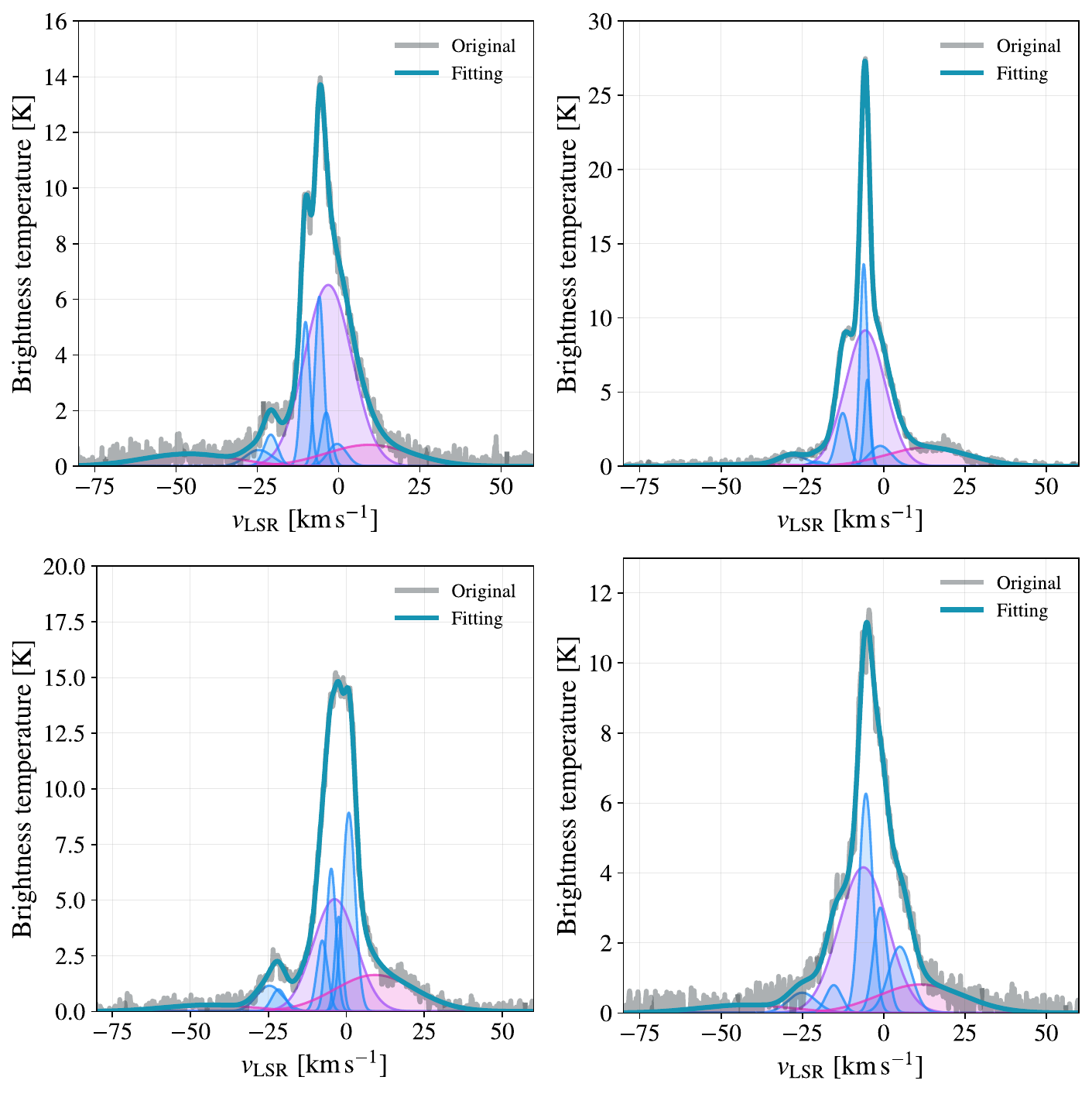} 
 \end{center}
\caption{Examples of ROHSA multi-Gaussian decomposition for four representative HI spectra. The gray lines show the observed spectra, while the red, purple, and blue curves represent the WNM, LNM, and CNM components, respectively. The green curves indicate the total fitted spectra.
{Alt text: Multi Gaussian fit to a neutral hydrogen spectrum. Brightness temperature is shown as a function of velocity with respect to the local standard of rest. Components correspond to cold, lukewarm, and warm neutral media, and their sum matches the observed profile.}}\label{fig:fig3}
\end{figure}

\begin{figure}
 \begin{center}
  \includegraphics[width=\columnwidth]{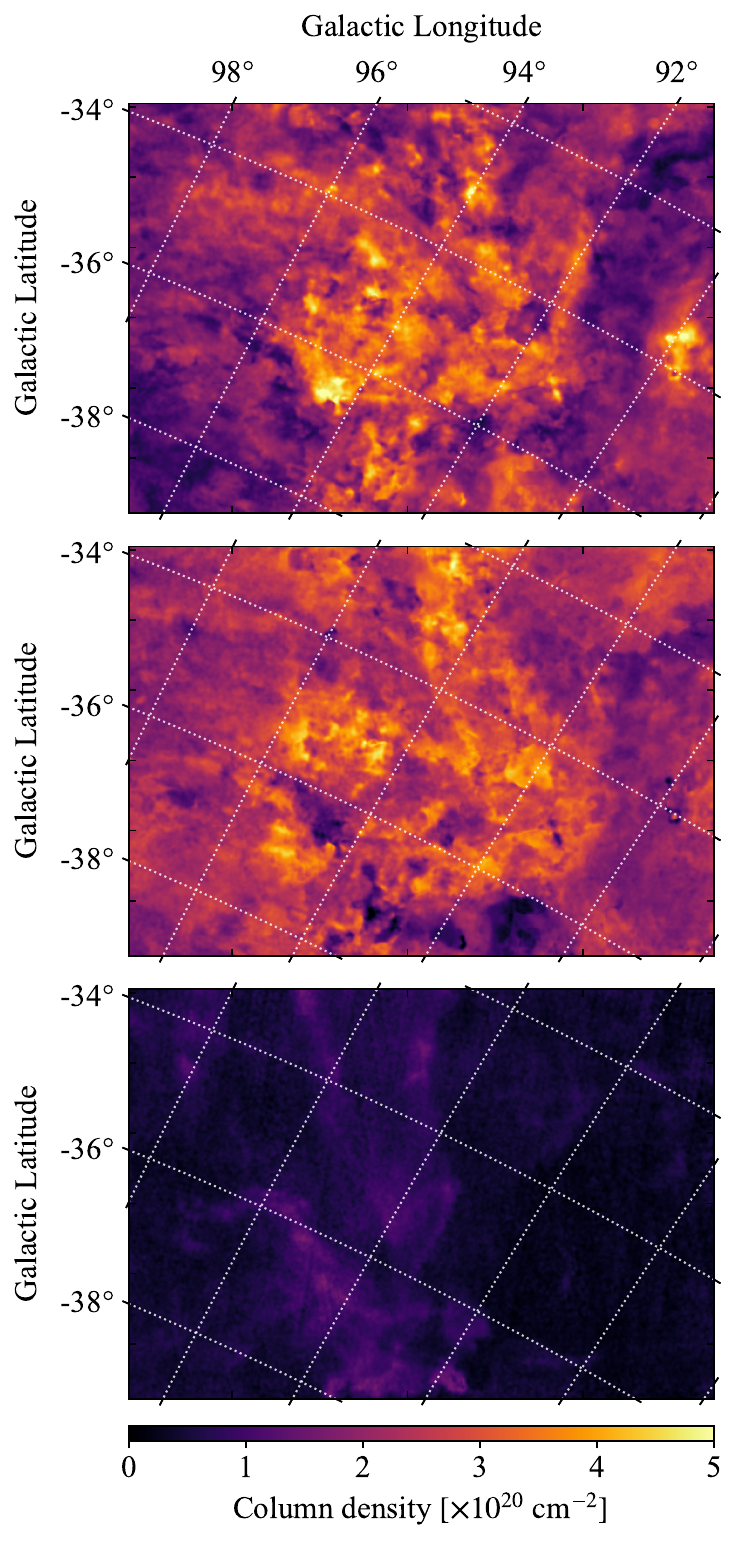}
 \end{center}
\caption{Column density maps of H\,\textsc{i} phases. (Top) CNM, (Middle) LNM, and (Bottom) WNM. The spatial domain is the same as in Figure \ref{fig:fig1}.
{Alt text: Three Galactic coordinate maps showing neutral hydrogen column density separated by phase. The top, middle, and bottom panels correspond to cold neutral medium, lukewarm neutral medium, and warm neutral medium, respectively.}}\label{fig:fig4}
\end{figure}

\begin{figure}
 \begin{center}
  \includegraphics[width=\columnwidth]{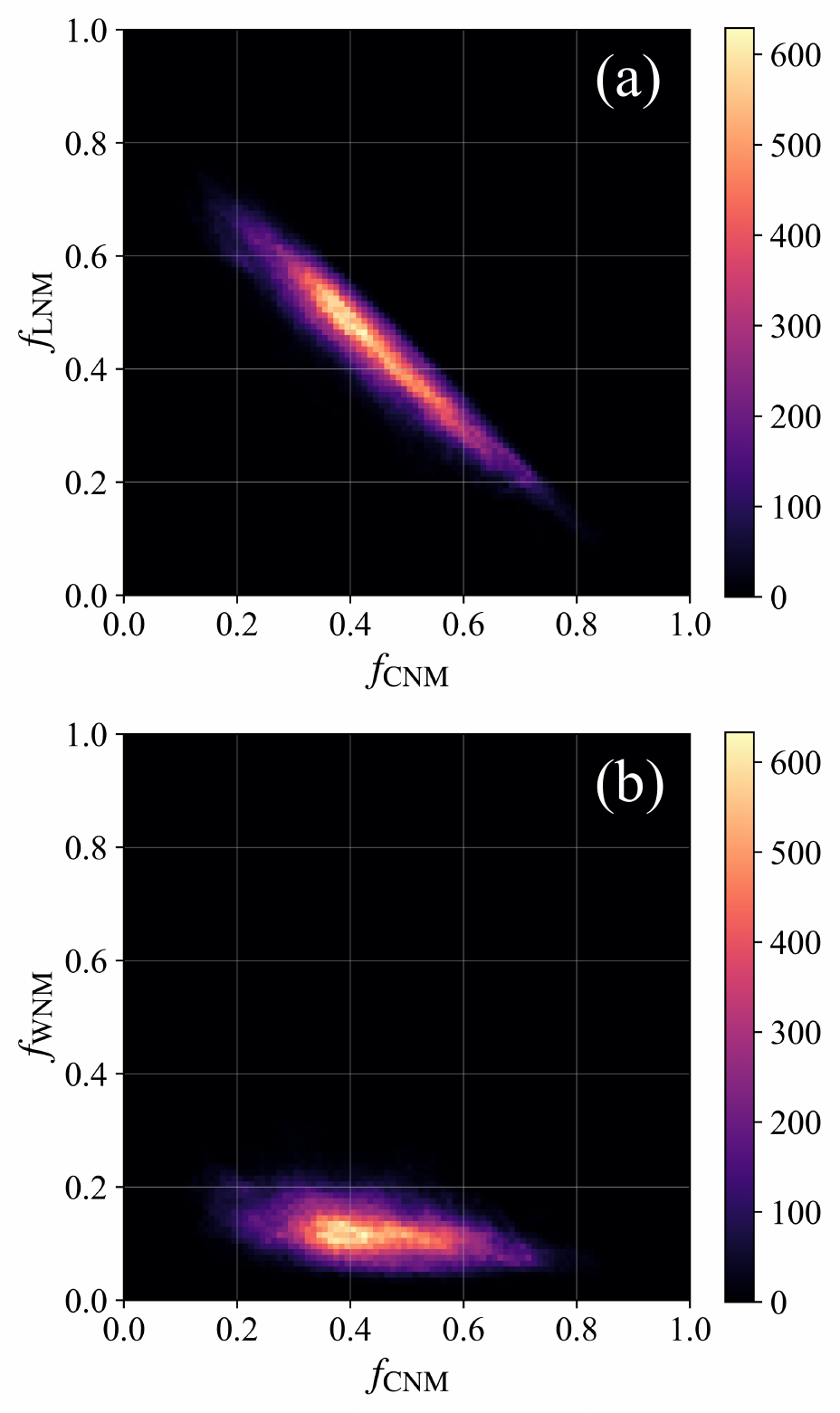}
 \end{center}
\caption{Two-dimensional frequency distribution of mass fraction in the HLCG 92$-$35 region. (a) Frequency distribution of $f_\mathrm{WNM}$ and $f_\mathrm{LNM}$, (b) Frequency distribution of $f_\mathrm{LNM}$ and $f_\mathrm{CNM}$.
{Alt text: Two vertically stacked density plots showing the frequency distribution of neutral hydrogen mass fractions in the HLCG 92$-$35 region. Panel a presents the mass fraction of warm neutral medium versus cold neutral medium. Panel b presents the mass fraction of lukewarm neutral medium versus cold neutral medium.}}\label{fig:fig5}
\end{figure}

\begin{figure}
 \begin{center}
  \includegraphics[width=\columnwidth]{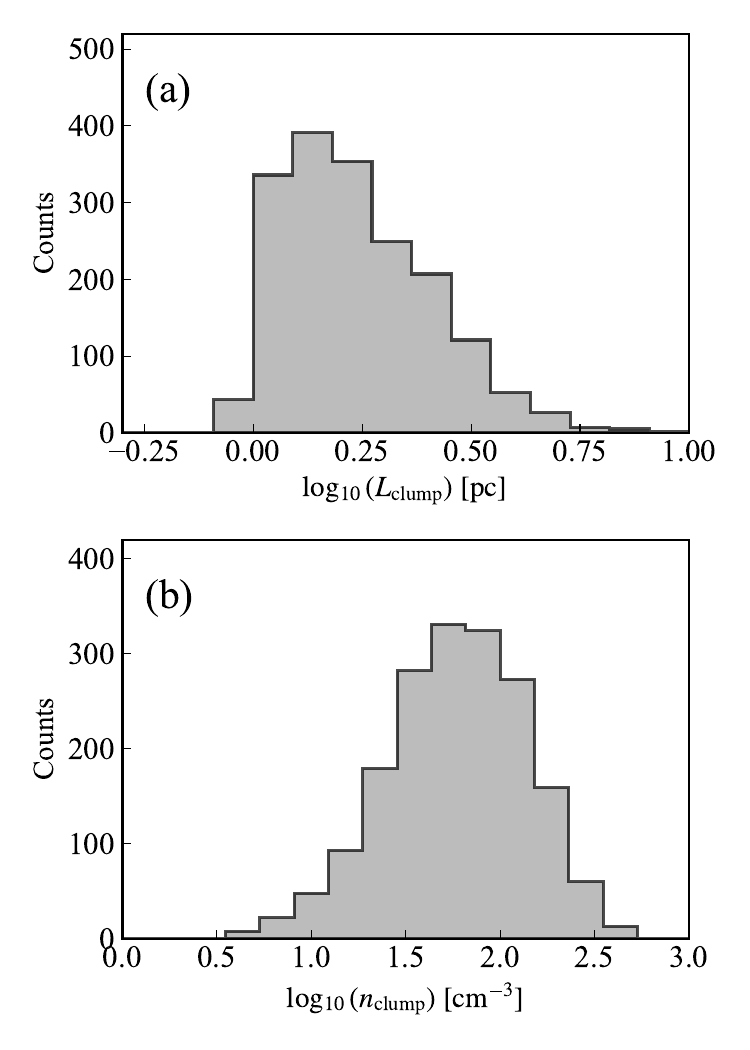} 
 \end{center}
\caption{(a) Size (diameter) histogram of CNM clumps. (b) Number density histogram CNM clumps.
{Alt text: Two histograms showing statistical distributions of cold neutral medium clumps. The upper panel presents the logarithm of effective radius in parsecs, and the lower panel shows the logarithm of number density in cubic centimeters. The vertical axis in both panels represents counts.}}\label{fig:fig6}
\end{figure}

\begin{figure}
 \begin{center}
  \includegraphics[width=\columnwidth]{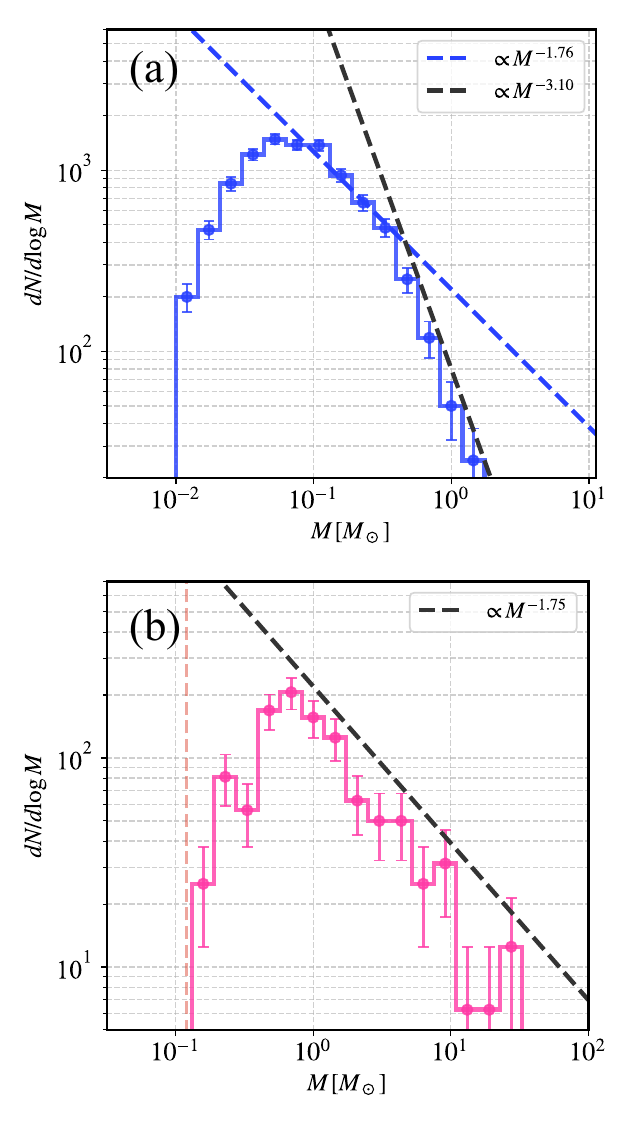}
 \end{center}
\caption{(a) CNM mass spectrum. The lines indicate reference slopes of $dN/dM \propto M^{-1.76}$ and $\propto M^{-3.11}$. The CNM clump distribution deviates from the $-1.76$ power law at both the low-mass and high-mass ends. (b) CO clump mass spectrum (red). The CNM clump mass spectrum from Figure \ref{fig:fig7}(a) is shown in blue for comparison. The solid line indicates a reference slope of $dN/dM \propto M^{-1.75}$, and the red dashed line represents the CO clump detection limit. 
{Alt text: Two panels displaying mass spectra of cold neutral medium and carbon monoxide clumps. The horizontal axis shows mass in solar masses, and the vertical axis shows number per logarithmic mass interval, both on logarithmic scales.}}\label{fig:fig7}
\end{figure}


\begin{figure}
 \begin{center}
  \includegraphics[width=\columnwidth]{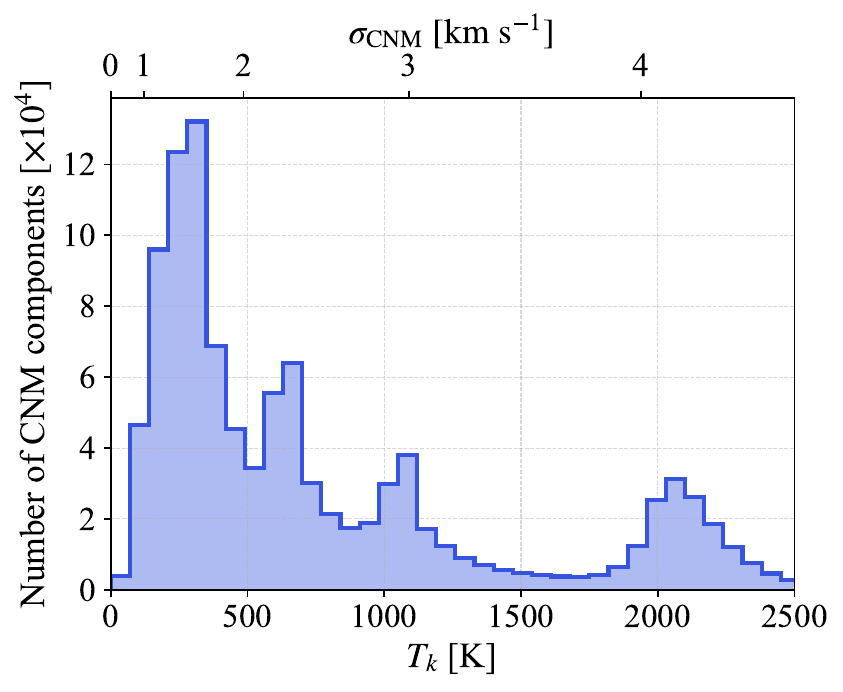} 
 \end{center}
\caption{Distribution of line widths for all CNM components separated by ROHSA, converted to temperature. The corresponding velocity dispersion ($\sigma$) is shown on the upper axis.
{Alt text: Histogram showing the distribution of line widths of cold neutral medium components separated by ROHSA. The lower axis gives temperature in kelvin converted from line width, and the upper axis shows the corresponding velocity dispersion in kilometers per second. The vertical axis represents the number of components.}}\label{fig:fig8}
\end{figure}

\begin{figure}
 \begin{center}
  \includegraphics[width=\columnwidth]{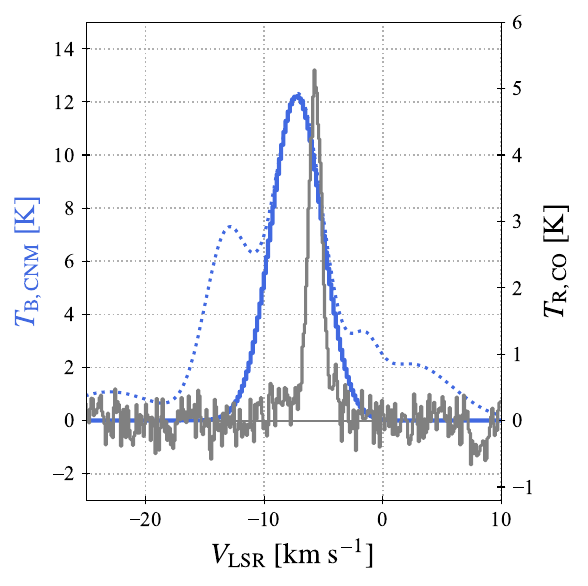} 
 \end{center}
\caption{Velocity offset at $(l, b) = (92.9^\circ, -34.7^\circ)$. The blue spectrum represents the CNM (dashed lines indicate all CNM components), and the gray spectrum represents $^{12}\mathrm{CO}(J=1\text{--}0)$.
{Alt text: Velocity spectra at Galactic longitude ninety two point nine degrees and latitude minus thirty four point seven degrees. The horizontal axis shows local standard of rest velocity in kilometers per second. The vertical axes show brightness temperature in kelvin for cold neutral medium and carbon monoxide emission. Two spectra are overlaid for comparison.}}\label{fig:fig9}
\end{figure}

\begin{figure}
 \begin{center}
  \includegraphics[width=\columnwidth]{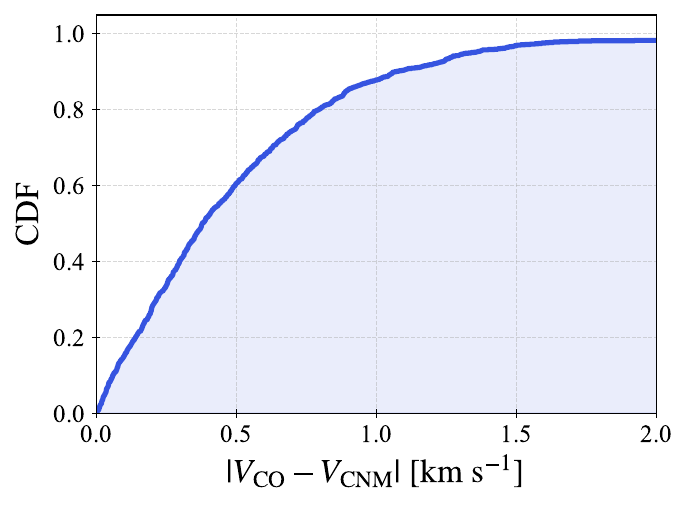} 
 \end{center}
\caption{Cumulative distribution function (CDF) of the absolute velocity difference between CNM and $^{12}$CO. The velocity difference is measured between each CO emission line and the CNM component closest to its central velocity. Only lines of sight where the CO peak intensity exceeds 2.0 K ($>4~\sigma$) are shown.
{Alt text: Cumulative distribution function of the absolute velocity difference between cold neutral medium and CO emission components. The horizontal axis shows velocity difference in kilometers per second, and the vertical axis shows cumulative fraction.}}\label{fig:fig10}
\end{figure}

\begin{figure}
 \begin{center}
  \includegraphics[width=\columnwidth]{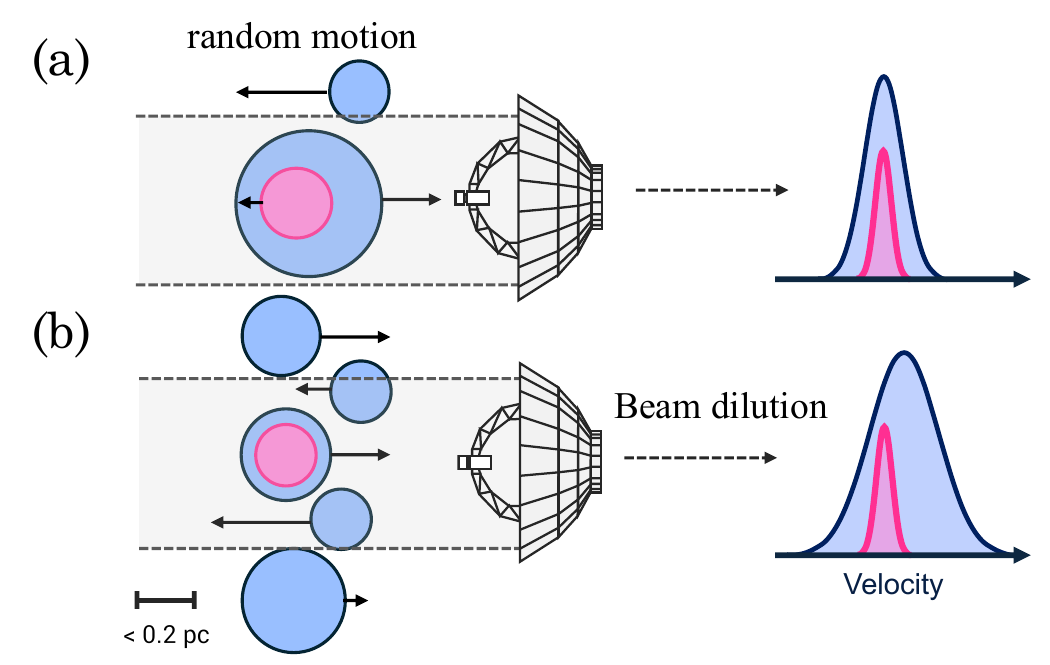} 
 \end{center}
\caption{Schematic illustration of a possible origin of the observed CNM--CO velocity offsets and linewidth broadening. (a) At the spatial resolution of the GALFA-H\,\textsc{i} observations, molecular clouds traced by $^{12}$CO($J=1$--0) appear to occupy only a limited velocity range within a single CNM component. (b) A possible underlying structure consisting of numerous unresolved CNM and molecular cloudlets. If molecular clouds form in only a subset of the CNM cloudlets, beam averaging can produce the apparent velocity offsets between CNM and CO emission seen in low-resolution observations. The velocity dispersion among unresolved CNM cloudlets may also broaden the observed CNM line profile.
{Alt text: Schematic comparison between low-resolution observations and an unresolved cloudlet model, showing how multiple CNM and molecular cloudlets can produce apparent CNM--CO velocity offsets and broadened CNM line profiles.
}}\label{fig:fig11}
\end{figure}

\begin{table*}[htbp]
\centering
\caption{Average peak intensity and velocity dispersion $\sigma$ of the nine components ($G_1-G_9$) separated by ROHSA in the HLCG 92$-$35 region.}
\begin{tabular}{c|ccccccccc}
\hline
Component 
& $G_1$ & $G_2$ & $G_3$ & $G_4$ & $G_5$ & $G_6$ & $G_7$ & $G_8$ & $G_9$ \\
\hline
peak intensity [K]
& 7.3 & 6.8 & 3.4 & 4.7 & 7.1 & 1.2 & 0.4 & 1.3 & 0.4 \\
\hline
velocity dispersion $\sigma$ [$\mathrm{km\,s^{-1}}$]
& 7.1 & 1.5 & 2.9 & 1.6 & 1.6 & 4.0 & 13.1 & 2.3 & 14.0 \\
\hline
\end{tabular}
\label{table:table2}
\end{table*}

\begin{figure*}
 \begin{center}
  \includegraphics[width=2\columnwidth]{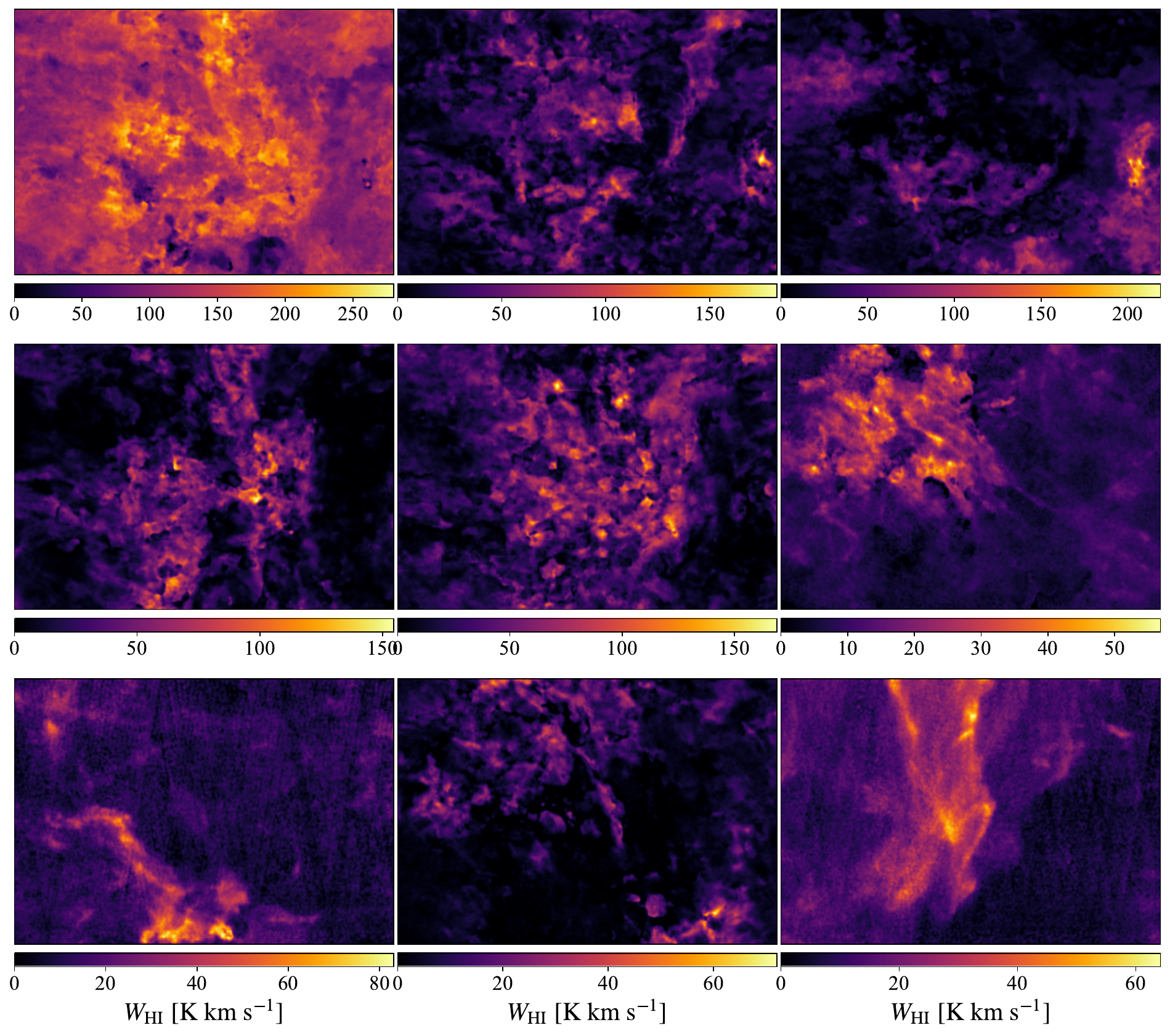} 
 \end{center}
\caption{Integral intensity maps of the nine H\,\textsc{i} components in the HLCG 92$-$35 region isolated by ROHSA (left column: $G_1, G_4, G_7$; center: $G_2, G_5, G_8$; right: $G_3, G_6, G_9$). The displayed region is the same as in Figure \ref{fig:fig1}. The color bar scale varies by map. Coordinates are the same as in Figure \ref{fig:fig1}.
{Alt text: Nine integrated intensity maps of neutral hydrogen components separated by ROHSA in the HLCG 92$-$35 region.}}\label{fig:fig13}
\end{figure*}

\begin{figure*}
 \begin{center}
  \includegraphics[width=\columnwidth]{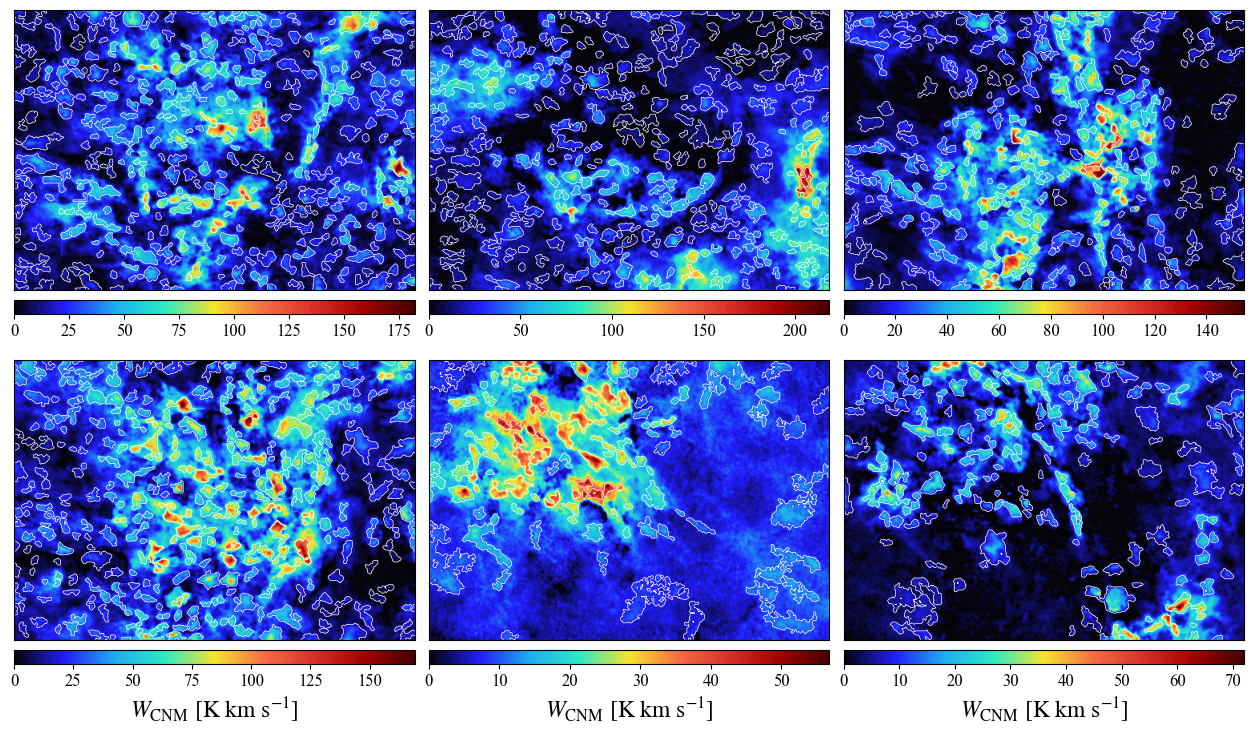} 
 \end{center}
\caption{CNM clumps in the HLCG 92$-$35 region identified with Astrodendro. Clump structures are shown as contours over the integrated intensity map of six CNM components. Left column: $G_2$ and $G_5$; center column: $G_3$ and $G_6$; right column: $G_4$ and $G_8$.
{Alt text: Spatial distribution of cold neutral medium clumps identified with Astrodendro in the HLCG 92$-$35 region.}}\label{fig:fig14}
\end{figure*}

\begin{figure}
 \begin{center}
  \includegraphics[width=\columnwidth]{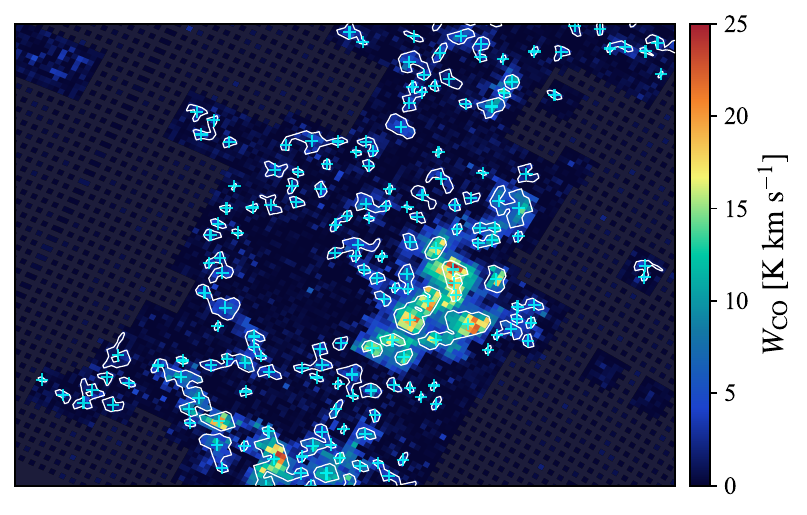} 
 \end{center}
\caption{CO clumps in the HLCG 92$-$35 region identified with Astrodendro. CO clump structures are overlaid as contours on the $^{12}\mathrm{CO}(J=1\text{--}0)$ integrated intensity map. The “+” symbols indicate the clump centroids. The displayed area corresponds to that in Figure \ref{fig:fig1}; clumps with centroids outside the analyzed region were excluded from the analysis.
{Alt text: Spatial distribution of CO clumps in the HLCG 92$-$35 region.}}\label{fig:fig15}
\end{figure}

\begin{figure*}[t]
 \begin{center}
  \includegraphics[width=0.9\textwidth]{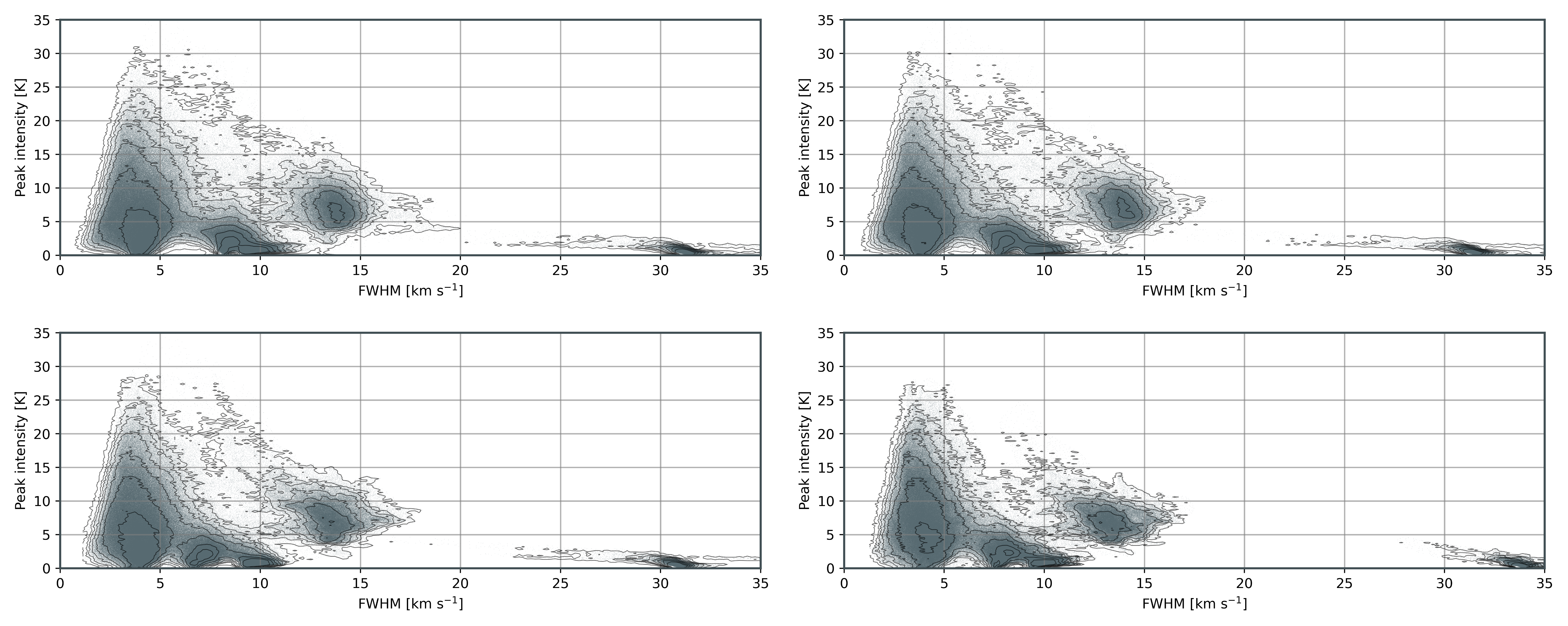} 
 \end{center}
\caption{FWHM--peak intensity distributions for $N=8$ and $\lambda'_\sigma=1$. 
The parameters $\lambda_a=\lambda_\mu=\lambda_\sigma$ are set to 1 (top left), 
10 (top right), 100 (bottom right), and 1000 (bottom left).
{Alt text: FWHM–peak intensity distributions for N equals 8 and lambda sigma prime equals 1 at lambda a, lambda mu, and lambda sigma equal to 1, 10, 100, and 1000.}}\label{fig:fig16}
\end{figure*}

\begin{figure*}[t]
 \begin{center}
  \includegraphics[width=0.9\textwidth]{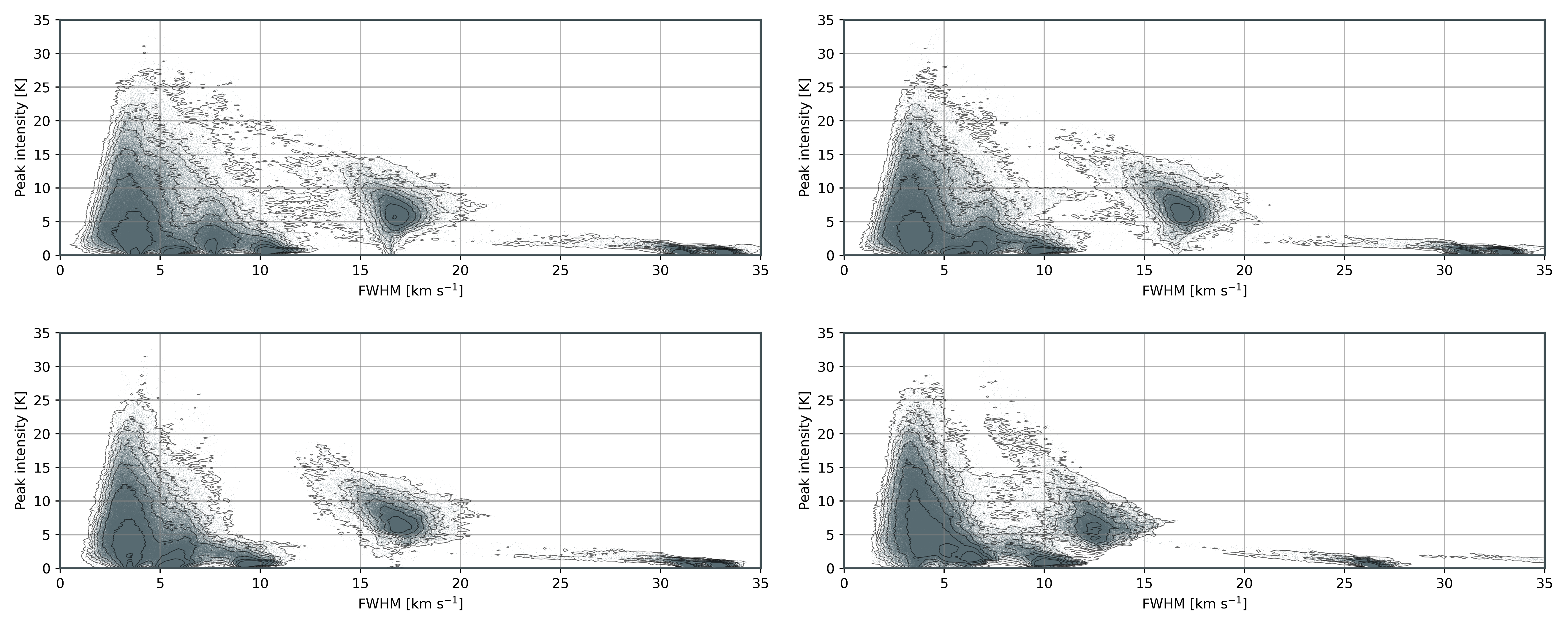} 
 \end{center}
\caption{Same as Figure~\ref{fig:fig16}, but for $N=9$.
{Alt text: FWHM–peak intensity distributions for N equals 9 and lambda sigma prime equals 1 at lambda a, lambda mu, and lambda sigma equal to 1, 10, 100, and 1000.}}\label{fig:fig17}
\end{figure*}

\begin{figure*}[t]
 \begin{center}
  \includegraphics[width=0.9\textwidth]{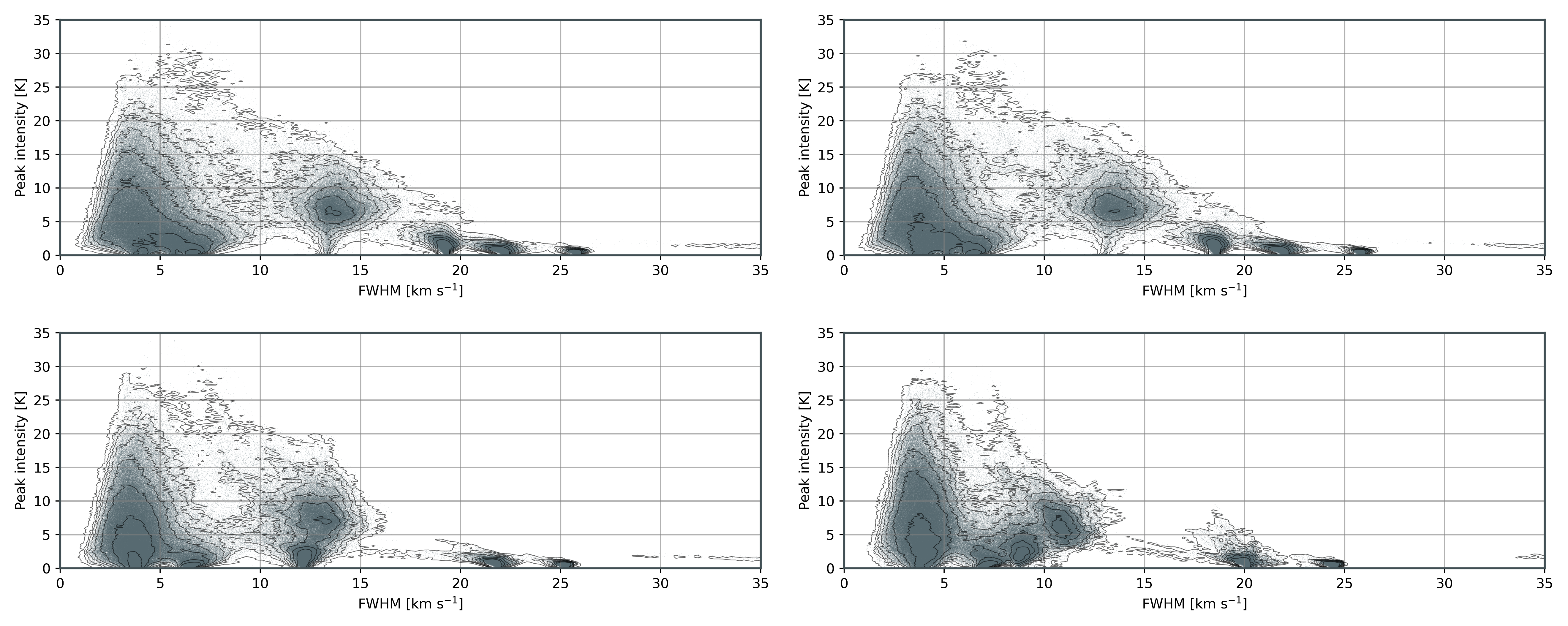} 
 \end{center}
\caption{Same as Figure~\ref{fig:fig16}, but for $N=10$.
{Alt text: FWHM–peak intensity distributions for N equals 10 and lambda sigma prime equals 1 at lambda a, lambda mu, and lambda sigma equal to 1, 10, 100, and 1000.}}\label{fig:fig18}
\end{figure*}

\begin{figure}
 \begin{center}
  \includegraphics[width=\columnwidth]{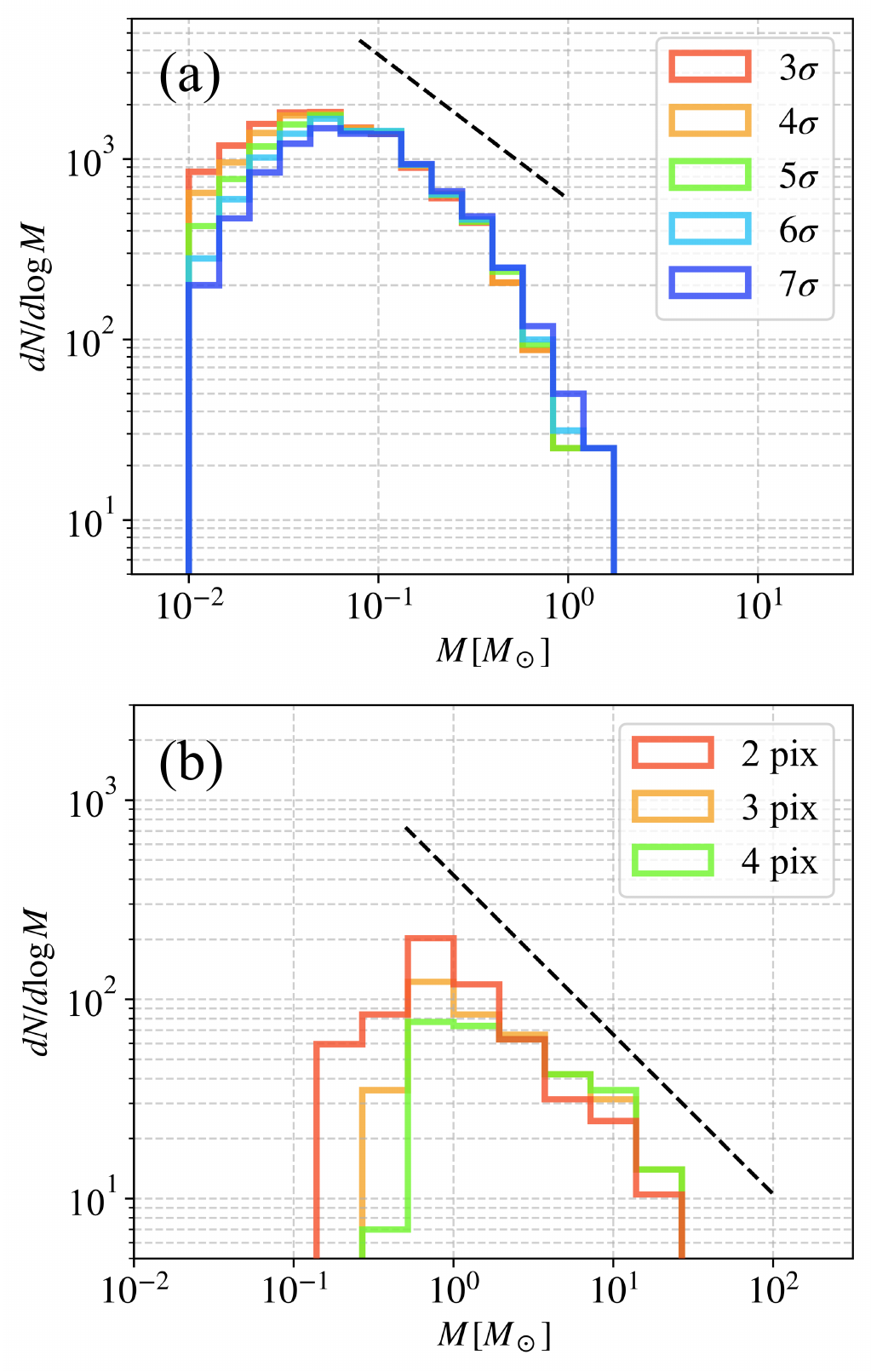} 
 \end{center}
\caption{
Mass spectra derived using different dendrogram parameters.
(a) CNM mass spectra for \texttt{min\_value} =
\texttt{min\_delta} = $3\sigma_\mathrm{rms}$--$7\sigma_\mathrm{rms}$
with \texttt{min\_npix} fixed at $25$.

(b) CO mass spectra for \texttt{min\_value} = $3\sigma_\mathrm{rms}$,
\texttt{min\_delta} = $2\sigma_\mathrm{rms}$,
and \texttt{min\_npix} = $2$--$4$ pixels.
In both panels, the black dashed line represents a power-law slope of $-1.8$.
{Alt text: Two-panel figure illustrating the robustness of the CNM and CO clump mass spectra against changes in dendrogram parameters. In the CNM analysis, increasing the detection threshold mainly suppresses the low-mass end of the spectrum, while the overall power-law slope remains similar. In the CO analysis, mass spectra obtained with smaller minimum clump-size thresholds are consistent with a power-law slope of approximately $-1.8$. When the minimum clump size is increased further, the number of identified clumps decreases substantially, leading to larger statistical fluctuations in the mass spectrum. Black dashed lines represent a power-law slope of $-1.8$.}}\label{fig:fig19}
\end{figure}

\end{document}